\documentclass[10pt,reprint,superscriptaddress,preprintnumbers,nofootinbib,amsmath,amssymb,aps,onecolumn,prd, floatfix ]{revtex4}

\usepackage{graphicx}% Include figure files
\usepackage{epsfig}
\usepackage{dcolumn}% Align table columns on decimal point
\usepackage{bm}% bold math
\usepackage{amssymb}
\usepackage{float}
\usepackage{amsmath}
\usepackage{subfigure}
\usepackage{cancel}
\usepackage[colorlinks]{hyperref}
\usepackage{array}
\usepackage[dvipsnames]{xcolor}
\usepackage{soul}
\hypersetup{
     breaklinks=true,
    pdfstartview={FitH},    % fits the width of the page to the window
    colorlinks=true,       % false: boxed links; true: colored links
    linkcolor=blue,          % color of internal links
    citecolor=red,        % color of links to bibliography
    filecolor=magenta,      % color of file links
    urlcolor=blue,           % color of external links
    anchorcolor=green,      % Color for anchor text
    linktocpage=true
}

\newcommand{\Mpl}{M_{\textrm{Pl}}}

\def\doi{http://doi.org}

\begin{document}

\title{Gravitational wave background from
quintessential inflation and NANOGrav data}

\author{Barnali Das}
\email{bd18ms201@iiserkol.ac.in}
\affiliation{Department of Physical Sciences,
Indian Institute of Science Education and Research Kolkata, Mohanpur-741 246, WB, India}
\author{Nur Jaman}
\email{nurjaman.pdf@iiserkol.ac.in} 
\affiliation{Department of Physical Sciences,
Indian Institute of Science Education and Research Kolkata, Mohanpur-741 246, WB, India}

\author{M Sami}
\email{ sami\_ccsp@sgtuniversity.org} 
\affiliation{SGT University, Gurugram, Delhi- NCR, Haryana- 122505, India}
\affiliation{Eurasian International Centre for Theoretical Physics, Astana, Kazakhstan}
\affiliation{Chinese Academy of Sciences,
52 Sanlihe Rd, Xicheng District, Beijing}
\begin{abstract}
We investigate the production process of induced gravity waves due to large scalar fluctuations in the paradigm of quintessential inflation. We numerically solve the Mukhanov-Sasaki equation for different sets of parameters to obtain the power spectra.
We demonstrate that the induced gravity wave signal generated in this framework can fall within the region of the NANOGrav data for chosen values of model parameters. We show that there is an allowed region of parameter space where the effect shifts to high frequency regime relevant to LISA and other available sensitivities.
\end{abstract}

\maketitle
\section{Introduction}
%%%%%%%%%%%%%%%%%%%
One of the generic features of cosmic inflation \cite{Guth:1980zm,starobinsky,Linde:1981mu}, apart from addressing the shortcomings of the hot big-bang model, is the production of primordial perturbations of quantum nature. Scalar fluctuations serve as the seed, necessary for structure formation, while the tensor part, known as Gravitational Waves (GWs), can be used as a tool to probe small scales. The detection of the relic GWs is regarded as a clean signal for confirming the paradigm of inflation.
%Large-scale perturbations, especially scalar perturbations, have been extensively studied and tightly constrained by the Cosmic Microwave Background(CMB) \cite{Planck:2018jri,Planck:2018vyg}. 
%Although CMB sets an upper bound for tensor perturbations, it constrains the ratio of tensor-to-scalar perturbations to be less than 0.05 ($r < 0.05$) at scale $k=0.05 \rm Mpc^{-1}$. 
%%%%
Recently, there has been a remarkable development in observational cosmology; different collaborations, including NANOGrav \cite{NANOGrav:2023gor,NANOGrav:2023hde}, European PTA (EPTA)/Indian PTA (InPTA) \cite{Antoniadis:2023rey, Antoniadis:2023bjw, Antoniadis:2023zhi}, Parkes PTA (PPTA) \cite{Zic:2023gta, Reardon:2023zen, Reardon:2023gzh}, and Chinese PTA (CPTA) \cite{Xu:2023wog}, have presented convincing evidence in support of Stochastic Gravitational Wave Background (SGWB) within the nHz frequency range. These findings are thought to be linked mainly to astrophysical sources. The standard interpretation is provided in terms of inspiraling super massive black hole binaries (SMBHBs). However, some alternative explanation seems to better fit the data as there is a deviation of the fit from SMBHB interpretation, especially in the high frequency tail (see Fig.~1 (a) of  \cite{NANOGrav:2023gor}). Hence, it is important to explore as an alternative, their cosmological origin. Indeed, the Induced Gravity Waves (IGWs) due to large scalar fluctuations generated during inflation could be a potential source for the observed SGWBs. This possibility has been discussed in NANOGrav investigations \cite{NANOGrav:2023hvm} and other subsequent works \cite{Franciolini:2023pbf,Inomata:2023zup,Cai:2023dls,Wang:2023ost,You:2023rmn,Yi:2023mbm,Liu:2023ymk,Balaji:2023ehk,Figueroa:2023zhu} (see also \cite{Bhattacharya:2020lhc}). In \cite{NANOGrav:2023hvm}, the statistical analyses seem to favor the IGWs \footnote{(in ref. \cite{NANOGrav:2023hvm} the termed SIGWs is equivalent to our term  IGWs)} with a high Bayes factor. The statistics are done with IGW + SMBHB and IGW alone. In this regard we expect the signal or part of it may be from IGW as we have done in our analysis. The high precision data will be relevant  to further clarify the issue.
%{\color{blue} The standard interpretation of these gravitational waves detected is provided in terms of inspiraling super massive binary black holes (SMBH). However, some alternative explanation seems to better fit the data. There is a deviation of the fit from SMBH interpretation, especially in the high frequency tail (see Fig. 1 (a) of  \cite{NANOGrav:2023gor}).  In the alternative (cosmological) interpretation of the data, the statistical  analysis seems to favor the IGWs \footnote{(in ref. \cite{NANOGrav:2023hvm}  the termed  SIGWs is equivalent to our term  IGWs)} with a high Bayes factor \cite{NANOGrav:2023hvm}. The statistics is done with  IGW + SMBH and IGW alone. In this regard we expect the signal or part of it  may be from IGW as we did in our analysis. The high precision data will be relevant  to further clarify the issue. }
 
In the context of standard cosmology, significant GW amplitude could be generated through two mechanisms: considering a stiffer epoch followed by inflation {\it \' a la} quintessential inflation, and secondly, sourcing the tensor modes from the scalar perturbations in the second order of perturbations. The second one is generic and applicable to any class of inflation, with some added features. Meanwhile, the first one holds true for a specific class of inflationary scenarios where one can generate a strong GW signal in the high frequency regime. 

The post-inflationary behaviour of the inflaton divides the inflationary models into two distinguished categories: (1) models in which inflation is followed by inflaton oscillations and its decay gives rise to (p)reheating \cite{Kofman:1994rk,Allahverdi:2010xz,Lozanov:2019jxc} ; (2) models with runaway type of potential where the field survives after inflation
to play an important role at late times, such as late time acceleration dubbed quintessential inflation \cite{Ratra:1987rm,Peebles:1998qn,Sahni:2001qp,Sami:2004ic,Peloso:1999dm,Peebles:1999fz,Copeland:2000hn,Rosenfeld:2005mt,Majumdar:2001mm,Dimopoulos:2000md,Hossain:2014zma,Sami:2004xk,WaliHossain:2014usl,Tashiro:2003qp,Dimopoulos:2017zvq,Dimopoulos:2001ix,Giovannini:2003jw,Tsujikawa:2013fta,Hossain:2014xha,Ahmad:2019jbm,deHaro:2021swo,Bettoni:2021qfs,Benisty:2020qta}. In the latter case, an alternative reheating mechanism such as instant preheating \cite{Felder:1998vq,Bassett:2005xm} is invoked: inflaton is coupled to a scalar field which in-turn is coupled to matter field such that a desired numerical value of reheating temperature 
is obtained by choosing appropriate numerical values of the couplings \cite{Ahmad:2019jbm}. The steep post-inflationary character of potential allows the commencement of radiative regime. In this case, inflation is followed by the kinetic regime, the duration of which depends upon the reheating temperature. The lower bound of this temperature is determined by the Big Bang nucleosynthesis (BBN) constraint due to presence of (primary) relic gravity waves \cite{Figueroa:2018twl,Kuroyanagi:2008ye,Ahmad:2017itq,Kuroyanagi:2014qaa}. A suitable type of steepness is asked for the scaling regime to be realized in radiation/matter era making the late time evolution independent of field initial conditions \cite{Ferreira:1997hj,Copeland:1997et,Steinhardt:1999nw}. The paradigm of quintessential inflation has a generic feature which distinguishes it from the standard scenario of inflation, namely, relic gravity waves have a blue tilted GW spectrum in high frequency region in this case.

Let us note that, in the linear order of perturbation theory, the tensor mode generated by quantum fluctuations evolves independently, known as the primary GWs mentioned so far, which is not true in the higher orders. For instance, in the second order of perturbation, the scalar and tensor modes couple together, and scalar fluctuations can source the secondary tensor modes \cite{Matarrese:1992rp, Matarrese:1993zf,Matarrese:1997ay,Noh:2003yg,Carbone:2004iv,Nakamura:2004rm}. These tensor  or IGWs are studied in \cite{Mollerach:2003nq,Ananda:2006af,Baumann:2007zm,Alabidi:2012ex,Alabidi:2013lya,Choudhury:2013woa,Kohri:2018awv,Domenech:2021ztg,Domenech:2019quo,Domenech:2020kqm,Correa:2023whf,Espinosa:2018eve,Ragavendra:2021qdu}. 
The secondary GWs dominate over the primary GWs for sufficiently large scalar perturbations on the relevant scales \cite{Ananda:2006af,Edgar:2010,Alabidi:2012ex}. The generation of large fluctuations in the early universe has been under great scrutiny recently. Apart from what we discussed above, they have other important implications for primordial black holes (PBHs) \cite{Hawking:1971ei, Carr:1974nx, HosseiniMansoori:2023mqh}, candidates for supermassive black holes present in our galaxies, which could unveil the secrets of nature related to the origin of dark matter \cite{Carr:2016drx} and baryon asymmetry \cite{Baumann:2007yr,Fujita:2014hha}. These fluctuations could be generated by the presence of a sharp transition from slow roll to ultra-slow regime due to a point of inflection or a bump/ dip in the inflationary potential.
In our work, we exploit the idea of introducing a bump \cite{Atal:2019cdz, Mishra:2019pzq} in the potential, effectively slowing down the rolling of the inflaton field giving rise to large scalar fluctuations consistent with the Cosmic Microwave Background (CMB) bounds. 
In this paper, we present a realistic description for IGWs in a particular class of inflation including full numerical description for generating large scalar fluctuations.
\iffalse Recent pulsar timing arrays (PTAs) experiments, including those at the North American Nanohertz Observatory for Gravitational Waves (NANOGrav) \cite{NANOGrav:2023gor, NANOGrav:2023hde, NANOGrav:2023ctt, NANOGrav:2023hvm, NANOGrav:2023hfp, NANOGrav:2023tcn, NANOGrav:2023pdq, NANOGrav:2023icp}, European PTA (EPTA) \cite{Antoniadis:2023utw, EuropeanPulsarTimingArray:2023qbc}, Indian PTA (InPTA) \cite{Antoniadis:2023jxr, Antoniadis:2023rey, Antoniadis:2023bjw, Antoniadis:2023zhi}, Parkes PTA (PPTA) \cite{Zic:2023gta, Reardon:2023zen, Reardon:2023gzh}, and Chinese PTA (CPTA) \cite{Xu:2023wog}, have provided compelling evidence for SGWB signals in the nHz frequency range, confirming the existence of genuine GWs. The pulsar timing technique relies on millisecond pulsars (MSPs), a type of neutron star, with exceptional rotational stability, where precise measurements of radio pulse times of arrival (TOAs) using maser clocks enable a phase-connected timing model to account for pulsar rotations. PTAs observe a population of the most stable MSPs. 

\fi
%The IGWs, dealt with in this article, may fall within the region of the recent NANOGrav data. 

The article is organised as follows: In Sec.~\ref{sec:inflation}, we visit the background dynamics of the inflation field along with scalar perturbations generated during inflation, setting the premise for our central goal. In Sec.~\ref{sec:PGW}, we show the calculations involved in obtaining primary GW. Sec.~\ref{Sec:IGW} covers discussion on IGW and comparison with present and future detection sensitivities. In Sec.~\ref{sec:conclusion}, we conclude our discussions.

\section{Inflation and Primary Power Spectrum} \label{sec:inflation}
In a spatially flat homogeneous and isotropic universe with FLRW metric $ds^2= -dt^2 + a^2(t)\delta_{ij}dx^i dx^j$, the inflationary dynamics or dynamics of the inflaton field $\phi$ are governed by the Friedmann and Klein-Gordan Eqs. 
\begin{eqnarray}
   && H^2 = \frac{1}{3\Mpl^2}\left[\frac{\dot{\phi}^2}{2} + V(\phi)
    \right]\, \, ,\\
    && \Ddot{\phi} + 3H\dot{\phi} + \frac{dV(\phi)}{d\phi} = 0.
\end{eqnarray}
where $a(t)$ is the scale factor, $t$ is the cosmic time, $H = \dot{a}/a$ is the Hubble parameter and $V(\phi)$ is the inflaton potential. For a homogeneous background, inflaton field is time dependent only, $\phi=\phi(t)$. 

To facilitate inflation, we require the field $\phi$ to roll slowly down the potential, such that potential energy dominates over the kinetic energy. Additionally, to have a sufficient duration of inflation, we require acceleration of the inflaton field to be negligible. These are called slow-roll conditions and are described in terms of two slow-roll parameters $\epsilon_1$ and $\epsilon_2$ such that 
\begin{eqnarray}
\epsilon_1 &\equiv&-\frac{\dot{H}}{H^2} \ll 1 \, , \\
\lvert\epsilon_2\rvert &\equiv& \big\lvert\frac{1}{H\epsilon_1}\frac{d\epsilon_1}{dt}\big\rvert \ll 1\,.
\end{eqnarray}
Duration of inflation are measured in terms of e-foldings, $N=\int^{t_f}_{t_{i}}Hdt'$, where $t_i$ denotes the time at the beginning of inflation and $t_f$ refers to the end of inflation. One typically requires $60-65$ e-folds of inflation to address issues associated with standard Big-Bang \cite{Baumann:2009ds}. Observational aspects associated with inflation are quantified in terms of scalar and tensor perturbations. We discuss tensor perturbations in more details in Sec.~\ref{sec:GW}. Here we mainly focus on its scalar counterpart. The scalar perturbations are quantified in terms of curvature power spectrum for a particular mode with wave number $k$ defined as \cite{Bassett:2005xm,Baumann:2009ds, Kinney2009tasi} 
\begin{eqnarray}
\label{eq:PS}
    \mathcal{P}_{S}(k) \equiv \frac{k^3}{2\pi^2}|\mathcal{R}_{k}|^2\bigg|_{k\ll a H}\, , 
\end{eqnarray}
where $ \mathcal{R}_k$ is the Fourier component of comoving curvature perturbation (see appendix \ref{appendix:A} for details). Under the slow-roll approximation one gets a analytical expression at horizon crossing for this \cite{Bassett:2005xm},

\begin{eqnarray}
\label{eq:PSSR}
    \mathcal{P}_{S}(k) \simeq \frac{1}{8\pi^2\epsilon_1}\frac{H^2}{\Mpl^2}\bigg|_{k= a H}  \,.
\end{eqnarray}
On a similar note, one can also quantify the tensor perturbations $h_{\textbf{k}}$ using the tensor power spectrum, discussed in details in Sec.~\ref{sec:GW}, as follows:
\begin{eqnarray}
    \mathcal{P}_T (k)\ \equiv \frac{k^3}{\pi^2} \sum_\lambda|h_\textbf{k}^{\lambda}|^2 \bigg|_{k\ll a H}\,.
\end{eqnarray}
One can then obtain the slow roll approximated analytical expression for the tensor power spectrum as follows:
\begin{equation}
\mathcal{P}_{T}(k) \simeq \frac{2}{\pi^2}\frac{H^2}{M_{\rm Pl}^2}\bigg|_{k=aH}\,.
\end{equation}
We can further define two more quantities of importance, the spectral index, $n_s$, and the tensor to scalar ratio, $r$ which have been tightly constrained by the Planck 2018 data \cite{Planck:2018jri}. The numerical expression followed by the slow roll approximated results for $n_s$ and $r$ are:
\begin{eqnarray}
    n_s &=& 1 + \left(\frac{d \ln \mathcal{P}_{S}}{d \ln k}\right) \simeq 1 - 4\epsilon_1 + 2\epsilon_2, \\
    r &=& \frac{\mathcal{P}_{T}}{\mathcal{P}_{S}} \simeq 16\epsilon_1
\end{eqnarray}

\begin{figure}
    \centering
    \includegraphics[scale = 0.3]{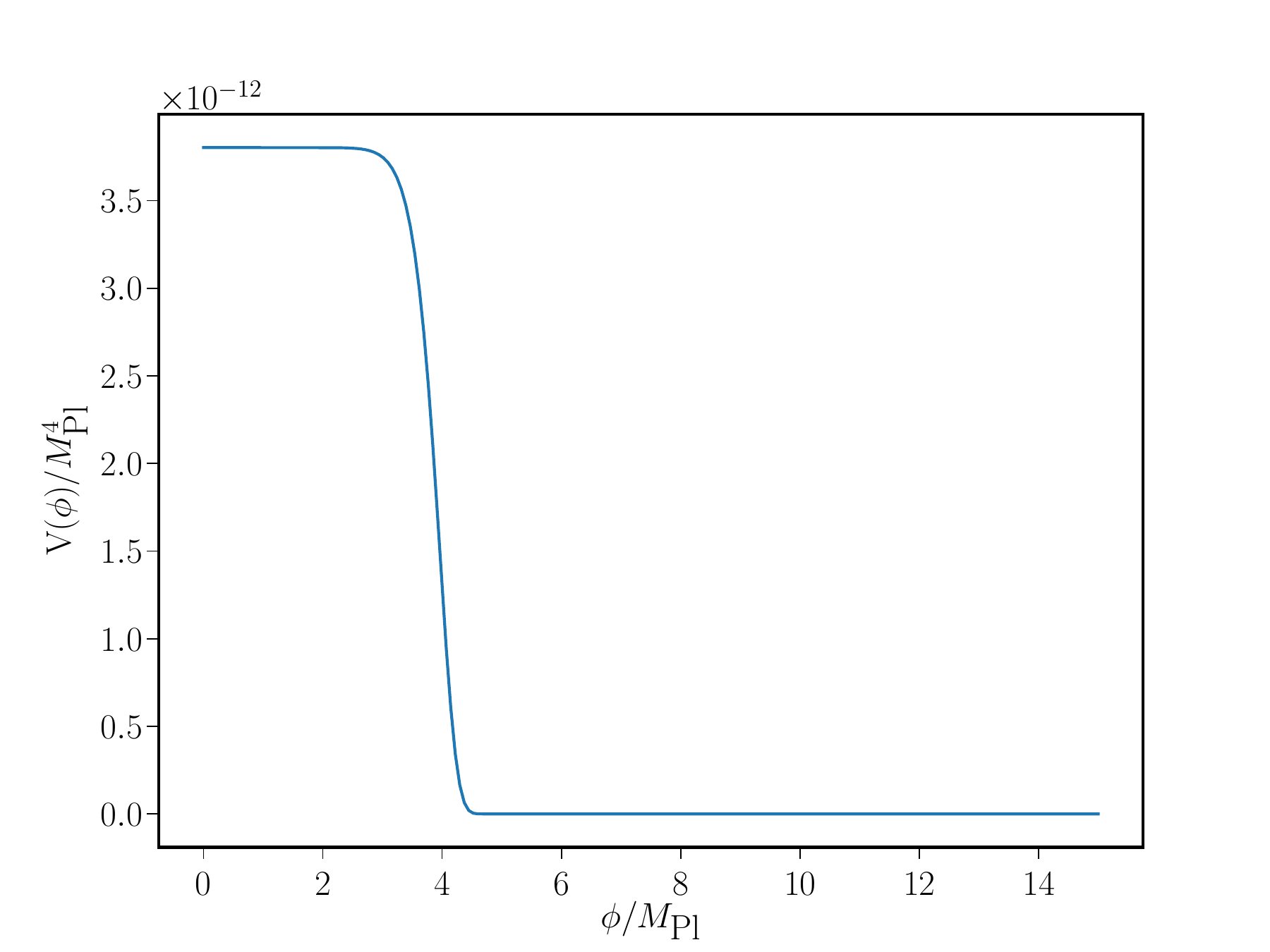}
    \caption{The figure depicts the typical quintessential potential. Initial flat region near the the small value of field accounts for inflation whereas the steep region governs the kinetic era followed by late time behaviour in the tail of the potential. Parameters choice are, $n= 15$ and $\lambda=10^{-9}$.}
    \label{fig:potential_quint}
\end{figure}

Setting the necessary ground, we now consider a particular quintessential inflationary potential introduced by \cite{Geng:2015fla} and further explored by\cite{Geng:2017mic,AresteSalo:2020yxl,Skugoreva:2019blk,Basak:2021cgk}:
\begin{eqnarray}
    V(\phi) = V_0\,\text{exp}\left[-\lambda \left(\frac{\phi}{\Mpl}\right)^n \right]
\end{eqnarray}
where $V_0$, $\lambda$ and $n$ are parameters. The slope and steepness are controlled by the set of $n$ and $\lambda$. The potential is shown in Fig.~\ref{fig:potential_quint}. 

It has been shown in literature \cite{Atal:2019cdz, Mishra:2019pzq}, that an addition of a bump or a dip to any inflationary potential can produce an inflection (like) point which decreases the speed of the inflaton field, eventually giving rise to a brief deviation from its slow-roll behaviour. Dropping in velocity can be be realized through the amplification in scalar power spectrum ($\mathcal{P}_S\propto \frac{1}{\epsilon_1} \sim \frac{H^2}{\dot \phi^2} $). In our case, we add a Gaussian bump of the form \cite{Mishra:2019pzq}:
\begin{eqnarray}
    f(\phi) = A\,\text{exp}\left[-\frac{1}{2}\frac{\left(\frac{\phi}{\Mpl}-\frac{\phi_0}{\Mpl}\right)^2}{\sigma^2}\right]
\end{eqnarray}
where $A, \sigma$ and $\phi_0$ are the height, width and position of the bump respectively.

After the inclusion of this bump, our full potential takes the form:
\begin{eqnarray}
    V(\phi) = V_0\,\text{exp}\left[-\lambda \left(\frac{\phi}{\Mpl}\right)^n \right]\left[1 + A\,\text{exp}\left(-\frac{1}{2}\frac{\left(\frac{\phi}{\Mpl}-\frac{\phi_0}{\Mpl}\right)^2}{\sigma^2}\right)\right]
\end{eqnarray}

Typically the parameter $V_0$ is fixed by the CMB normalization. So, the others parameters ($\lambda, n, A, \sigma$ and $\phi_0$) can be tuned within the limit from different constraints. 

\begin{figure}
    \centering
    \includegraphics[scale = 0.3]{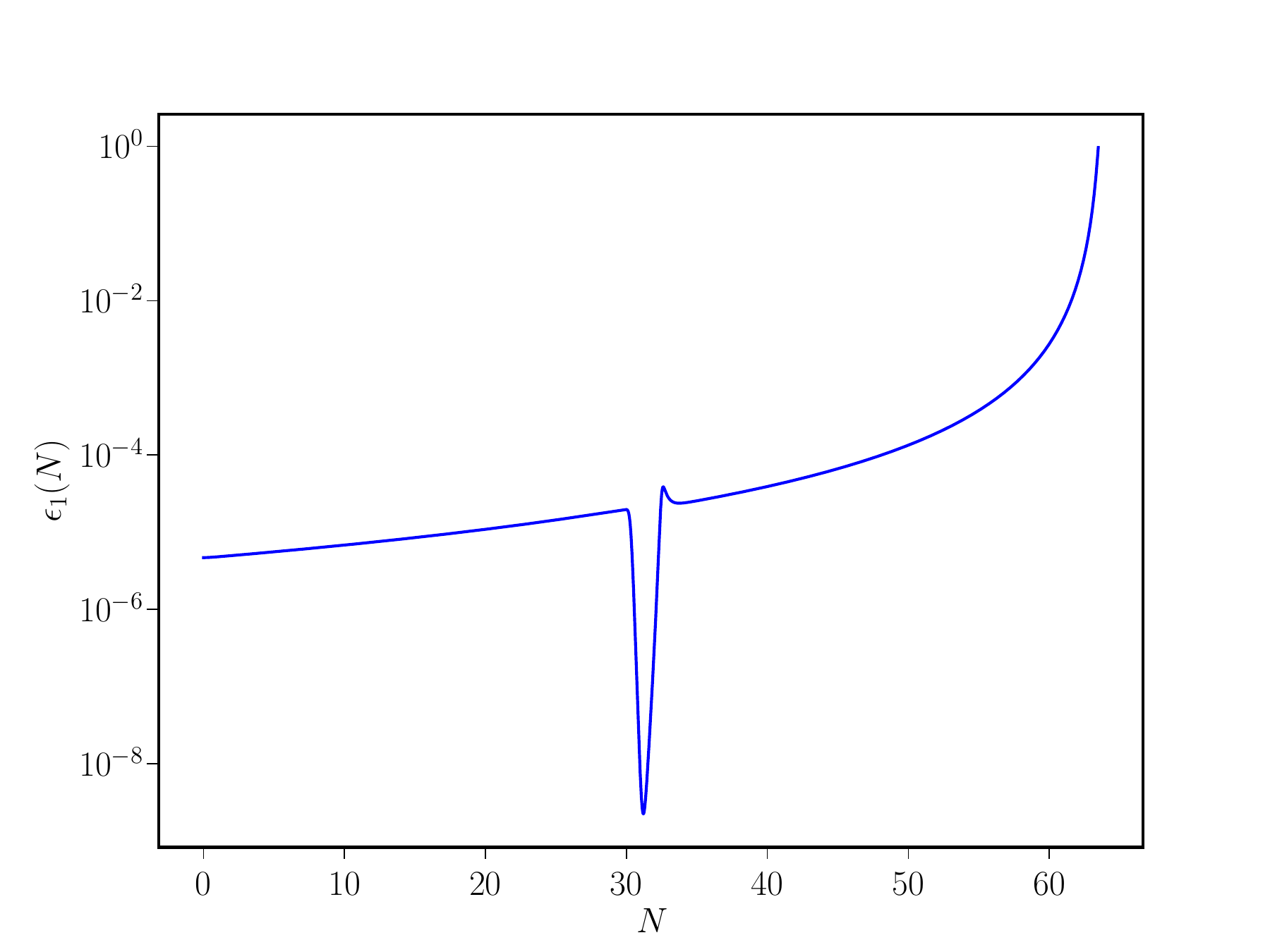}
    \hskip -0.2in
    \includegraphics[scale = 0.3]{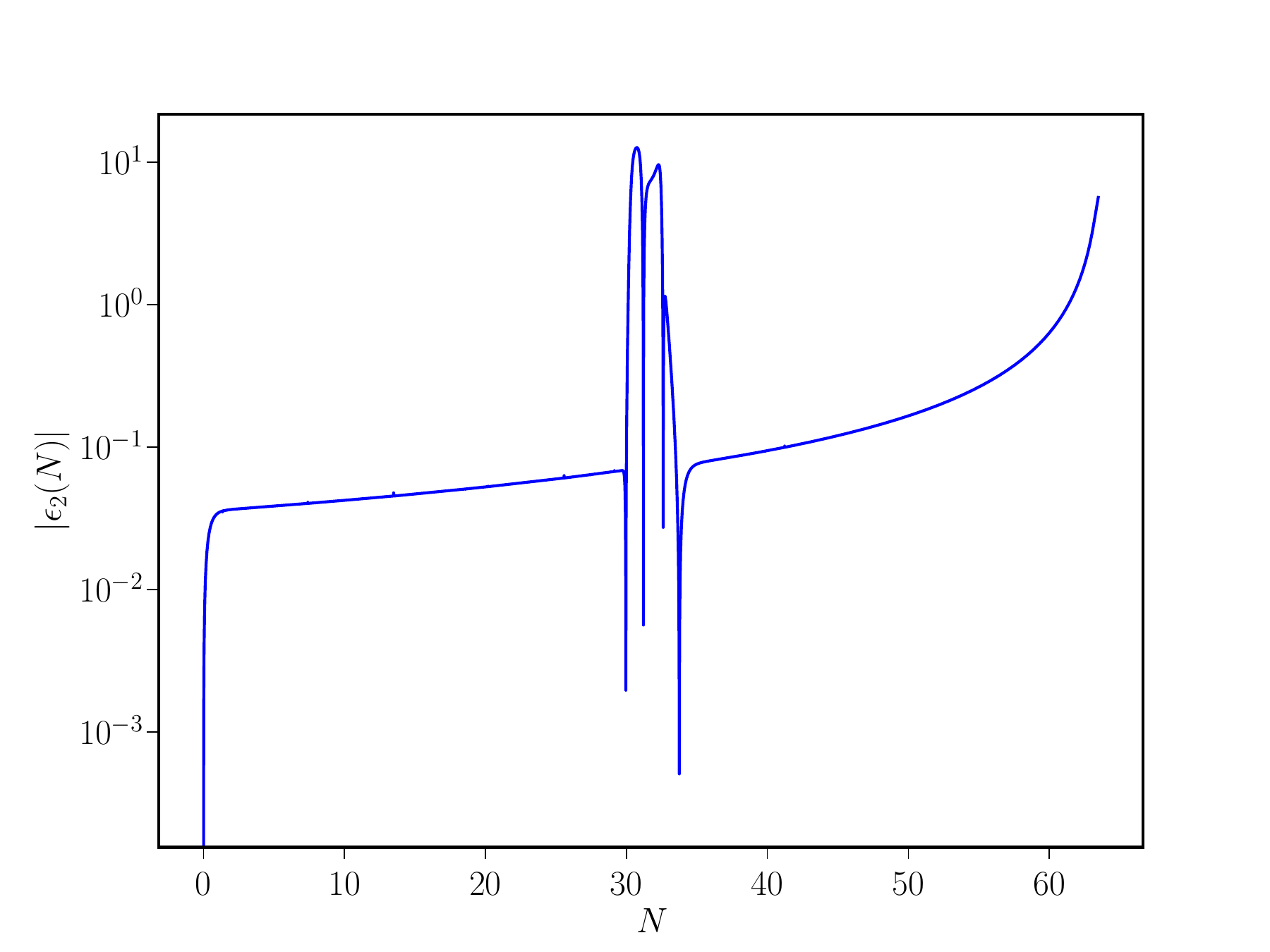}
    \caption{Behaviour of first (left) and second (right) slow-roll parameters. In the plot at right, $\epsilon_2$ deviates from the slow-roll behaviour around $\phi = \phi_0$, which occurs at $N \simeq 31$. This is for set I of parameter sets given in table.~\ref{table:parameters}.}
    \label{fig:epsilon}
\end{figure}

\begin{figure}
    \centering
    \includegraphics[scale = 0.4]{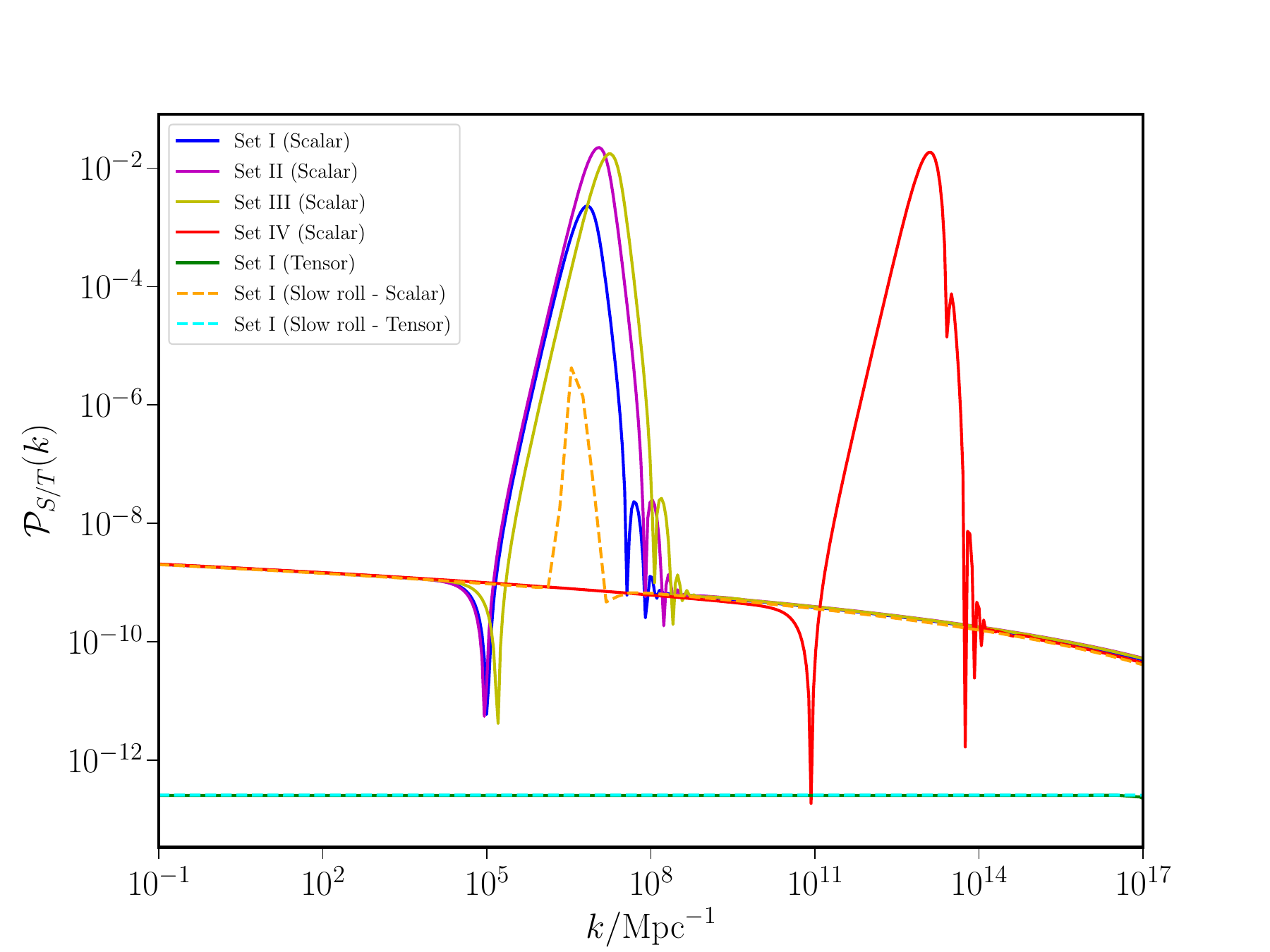}
    \caption{The figure shows the numerical power spectra for the four parameter sets as given in table:~\ref{table:parameters}. It also shows the tensor power spectrum and slow-roll approximated scalar and tensor power spectra for set I. The numerical and analytical results are comparable except at $k/\rm Mpc^{-1} \simeq 10^7$ and $k/{\rm Mpc}^{-1}\simeq 10^{13}$, which are due to the inclusion of bump. Different position of enhancement corresponds to different locations of the bump. }
    \label{fig:powerspectra}
\end{figure}

The complexity of the potential, demand a full numerical solution even for the Klein-Gordon equation for its background evolution. After solving the background dynamics, we clearly see that the second slow-roll parameter $\epsilon_2$, violates the slow-roll condition for a brief period as shown in Fig.~\ref{fig:epsilon}. This necessitates the numerical solution for perturbation equations, namely the Mukhanov-Sasaki equation (appendix~\ref{appendix:A}). We have solved the perturbation equations, for four different sets of parameters with values given in table.~\ref{table:parameters}. As we can see in Fig.~\ref{fig:powerspectra} the the power spectra are enhanced compared to its nearly scale invariant version near particular $k$ values. These corresponds to different $\phi_0$ values or the position of the bump. One should note that, slow-roll approximated solution for the power spectra shows a much less enhancement in its amplitude (see Fig.~\ref{fig:powerspectra}) \footnote{The background evolution and power spectra, along with similar plots, can be reproduced using the codes accessible at \href{https://github.com/mimibarnali00/Cosmology}{https://github.com/mimibarnali00/Cosmology}.}. These enhanced scalar fluctuations or power spectra act as a source for the IGWs of our interest discussed in Section~\ref{Sec:IGW}. In our analysis, the spectral index and tensor-to-scalar ratio ($n_s -r$) are always within the 2-sigma region given by Planck mission 2018 \cite{Planck:2018jri}.

\begin{table}
\centering
\begin{tabular}{ | m{2cm} | m{2cm}| m{3cm} | m{2cm} | } 
  \hline
  Sets & $\phi_0/\Mpl$ & $A$ & $\sigma$\\ 
  \hline
  \textcolor{blue}{Set I} & 2.527 & $1.3350 \times 10^{-5}$ & $10^{-3}$\\ 
  \hline
  \textcolor{magenta}{Set II} & 2.53 & $1.3680 \times 10^{-5}$ & $10^{-3}$\\ 
  \hline
  \textcolor{olive}{Set III} & 2.533 & $1.4011 \times 10^{-5}$ & $10^{-3}$\\ 
  \hline
  \textcolor{red}{Set IV} & 2.65 & $3.7220 \times 10^{-5}$ & $10^{-3}$\\ 
  \hline
\end{tabular}
\caption{The table displays four sets of parameter values that were used to obtain our results. All four sets share common values for potential parameters, specifically $V_0 = 3.8025 \times 10^{-12}$, $\lambda = 10^{-9}$, and $n = 15$. The parameters $\phi_0$ and $A$, which determine the position and shape of the potential, are varied to obtain different results. For consistency, we use color codes to represent each set: Set I - Blue, Set II - Magenta, Set III - Yellow, and Set IV - Red.}
\label{table:parameters}
\end{table}

%%%%%%%%%%%%%%%%%%%%%%%%%%%%%%%%%%%%%%%%%%%%%%%%%%%%%%%%%%%%%%%%%%%
\section{Gravitational Waves Backgrounds} \label{sec:GW}
\subsection{Primary Gravitational Waves} \label{sec:PGW}
The detection of primordial GWs background would be the smoking gun for justifying paradigm of inflation. They will provide insight into the underlying physics of the early Universe. For the case quintessential inflationary scenario, the presence of kinetic epoch followed by inflation blue tilts the primordial GW spectra at high frequency range \cite{Sahni:2001qp,Giovannini:1998bp,Giovannini:1999qj,Giovannini:1999bh,Riazuelo:2000fc,Giovannini:2008tm,Artymowski:2017pua,Figueroa:2018twl}. This feature in the GW background, makes the quintessential inflationary scenario distinguishable from the other. Some future detectors operating in range of high frequency can be relevant in this context \cite{Ahmad:2019jbm}.

Gravitational waves can be described as a transverse-trace-less part of the metric perturbation. At linear order perturbation theory, scalar, vector and tensor modes do not couple with each other. So, for primary gravitational waves in a spatially flat FLRW background, we can write the metric element as 
 \begin{eqnarray}
 ds^2 =-dt^2+a^2(t)(\delta_{ij}+h_{ij})dx^idx^j
\label{eq:tensormetric}
\end{eqnarray}
with the tensor mode satisfying $h_{ii} =h_{0 0}=\partial^i h_{ij}=0$. 
One can decompose $h_{ij} (t, x)$ into its Fourier mode with two polarization tensors, 
\begin{equation}
h_{ij}(t,\textbf{x})=\sum_{\lambda=+,\times}^{}\int\frac{d^3k}{(2\pi)^{3/2}}\epsilon_{ij}^{\lambda}
(\textbf{k})h_\textbf{k}^{\lambda}(t,\textbf{k})e^{i\textbf{k}\cdot\textbf{x}},
\end{equation}
where the polarization tensors $\epsilon_{ij}^{+,\times}$ satisfy symmetric and transverse-trace-less condition and are normalized as
$\sum_{i,j}\epsilon_{ij}^{\lambda}(\epsilon_{ij}^{\lambda^{\prime}})^*=2\delta^{\lambda\lambda^{\prime}}$. The Fourier modes satisfy the equation of motion
\begin{equation}
h^{\lambda ''}_\textbf{k} (\eta) + 2 \mathcal{H} h^{\lambda '}_\textbf{k} (\eta) + k^2 h_\textbf{k}^{\lambda} (\eta) = 0,
\label{eq:evoh}
\end{equation}
where `$'$' denotes derivative with respect to conformal time, $\eta, (d\eta=\frac{dt}{a}$) and $\mathcal{H}= \frac{a'}{a}$. The normalized GW energy density spectrum is defined by its energy density per logarithmic frequency, 
\begin{equation}
\Omega_{\rm GW} (k, \eta) \equiv \frac{1}{\rho_{\rm crit} (\eta)} 
\frac{d \rho_{\rm GW}}{ d \ln k },
\label{omegagw}
\end{equation}
where $\rho_{\rm crit}(\eta)$ is total energy density at conformal time $\eta$. The GW energy density $\rho_{\rm GW}$ is given by ($0,0$) component of the energy momentum tensor, 
\begin{eqnarray}
\rho_{\rm GW} = - T^{0}_{0} 
&=& \frac{\Mpl^2}{8} 
\frac{{\left( h'_{ij} \right)}^2 + {\left( \nabla h_{ij}\right)}^2}{a^2},\nonumber \\
&=& \frac{\Mpl^2}{4} \frac{{\left( \nabla h_{ij}\right)}^2}{a^2},\nonumber \\
& = & \frac{\Mpl^2}{4} \int \frac{d^3 k}{(2\pi)^3}\frac{k^2}{a^2} 2 \sum_\lambda|h_k^\lambda|^2.
\label{rhogw}
\end{eqnarray}
Now, the tensor power spectrum is defined as $\mathcal{P}_h (k)\ \equiv \frac{k^3}{\pi^2} \sum_\lambda|h_\textbf{k}^{\lambda}|^2 $. Thus Eq.~\eqref{omegagw} and Eq.~\eqref{rhogw} give
\begin{equation}
\Omega_{\rm GW,0} (k)=\frac{1}{12}\left(\frac{k^2}{a_0^2 H_0^2}\right)\mathcal{P}_h (k) =\frac{1}{12}\left(\frac{k^2}{a_0^2 H_0^2}\right) \mathcal{P}_{T}(k) T^2(k) \,.
\label{eq:GWprimary}
\end{equation}
 In the last step, we divide the power spectrum into primordial spectrum $\mathcal{P}_{T}(k)$ and transfer function $T^2 (k)$. The primordial tensor power spectrum at the horizon exit (under slow-roll approximation) is given by \cite{Bassett:2005xm}
\begin{equation}
\mathcal{P}_{T}(k)
= \frac{2}{\pi^2}\frac{H^2}{M_{\rm Pl}^2}\bigg|_{k=aH}.
\label{eq:powerprimary}
\end{equation}
The amplitude of the tensor fluctuation stays constant in the super Hubble region ($k\ll \mathcal{H}$) and decays as $ h_{k} ^\lambda \propto \frac{1}{a}$ inside the horizon ($k\gg \mathcal{H}$). The transfer Function $T^2(k)$ takes this decay into account \cite{Kuroyanagi:2008ye, Ahmad:2019jbm} and is evaluated to be 

\begin{equation}
T^2(k) = \frac{1}{2}\frac{a_{\rm hc}^2}{a_0^2},
\end{equation}
where $a_{\rm hc}$ and $a_0$ are the scale factors at horizon crossing and at present respectively. The expansion rate of the universe is given by Hubble parameter
\begin{equation}
H=  H_0\sqrt{
\Omega_\phi(a)+
\left(\frac{g_*}{g_{*0}}\right)
\left(\frac{g_{*s}}{g_{*s0}}\right)^{-4/3} 
\Omega_{\rm r0}\left(\frac{a}{a_0}\right)^{-4}
+\Omega_{\rm m0}\left(\frac{a}{a_0}\right)^{-3}
+\Omega_{\rm \Lambda 0}}\, , 
\label{eq:Hubble}
\end{equation}
where $\Omega_{\phi}$ is due to the presence of quintessence domination right after inflation, namely the kinetic epoch. The present dark energy domination is simply assumed to be the cosmological constant. In Eq.~\eqref{eq:Hubble}, $g_*$ and $g_{*s}$ are relativistic degrees of freedom for the energy density and entropy density respectively at temperature $T$. Their respective values at present are $g_{*0} = 3.36$ and $g_{*s0} = 3.91$. 
Considering the scale at horizon re-entry $k= a_{\rm hc} H_{\rm hc}$, and Hubble parameter at different epochs from Eq.~\eqref{eq:Hubble}, we calculate the transfer function. We then combine it with Eq.~\eqref{eq:GWprimary} and Eq.~\eqref{eq:powerprimary} to get the primary GW spectrum at today (for details see, \cite{Ahmad:2019jbm, Jaman:2022bho}) for mode re-entry during matter (M), radiation (R) and kinetic (K) era respectively as 
\begin{eqnarray}
\Omega_{\rm GW, 0}^{\rm (M)}
&=& \frac{1}{24}\Omega_{\rm m0}^2\frac{a_0^2 H_0^2}{k^2}
 \mathcal{P}_{T}
~~~~~~~~~~ 
(k_{\rm 0}<k \leq k_{\rm eq}),
\\
\Omega_{\rm GW, 0}^{\rm (R)}
&=& \frac{1}{24}\Omega_{\rm r0}\left(\frac{g_*}{g_{*0}}\right)
\left(\frac{g_{*s}}{g_{*s0}}\right)^{-4/3}  \mathcal{P}_{T}
~~~~~~~~~~  
(k_{\rm eq}<k \leq k_{\rm r}),
\label{eq:OGWRD}
\\
\Omega_{\rm GW, 0}^{\rm (K)}
&=& \Omega_{\rm GW, 0}^{\rm (R)}\left(\frac{k}{k_r}\right)
~~~~~~~~~~  
(k_{\rm r}<k \leq k_{\rm max}),
\label{eq:OGWKD}
\end{eqnarray}
where $k_0, k_{\rm eq}, k_{r}$ and $k_{\rm max} $ are the modes which re-enter at present, at matter radiation equality, at beginning of radiation era and at the end of inflation \footnote{ We have assumed the kinetic epoch followed just after end of inflation and present universe is approximated by matter domination, since the contribution for dark energy is not relevant for us.}. Considering the transition frequency
\begin{eqnarray}
    f = \frac{k}{2\pi} = 1.5 \times 10^{-15} \left(\frac{k}{\rm Mpc^{-1}}\right) \text{Hz}\,.
    \label{eq:freq_k}
\end{eqnarray}
at corresponding eras, we obtain the GW spectrum denoted by the green curve of Fig.~\ref{fig:GW01}. Also, BBN puts constraint on the maximum amplitude of GWs on relevant scale \cite{Figueroa:2018twl,Cyburt:2015mya,Jaman:2022bho}, 
 \begin{equation}
\label{BBNB}
\Omega_{\rm GW, 0}h^2<1.12\times 10^{-6},
\end{equation}

As we can see from Eq.~\eqref{eq:OGWKD} and from Fig.~\ref{fig:GW01}, the primary spectra is blue tilted in the kinetic regime. So, the maximum contribution relevant to BBN will come from this; thus, we have the following bound, 
\begin{eqnarray}
\Omega_{\rm GW, 0}^{\rm max} \approx \Omega_{\rm GW,0}^{(\rm K) \rm max}< 1.12\times 10^{-6}\,.
\label{BBNboundKin}
\end{eqnarray}
The last bound enables us to constrain the duration of kinetic era. For the case of instant preheating scenario, this bound translates into \cite{Jaman:2022bho}
\begin{eqnarray}
\frac{T_{\rm end}}{T_r}\sim \frac{a_{\rm end}}{a_r}< 5.5 \times 10^4\,.
\end{eqnarray}
This condition allows us to control the length of the kinetic regime, which spans from the conclusion of inflation to the onset of the radiation era. If the energy density of radiation generated at the end of inflation is higher, the duration of the kinetic regime will be shorter, making it easier to satisfy the BBN bound. 

\subsection{Induced Gravitational Waves} \label{Sec:IGW}
For modelling IGWs, we first need to consider a metric having both scalar and tensor perturbations, which, in Newtonian gauge, in the absence of anisotropy, can be written as \cite{Kohri:2018awv, Baumann:2007zm}:
\begin{eqnarray}
    ds^2 = a(\eta)^2\biggl[ -\left(1+2 \Phi\right)d\eta^2 +\left(\left(1-2 \Phi\right)\delta_{ij}+ \frac{h_{ij}}{2}\right)dx^i dx^j
\biggr] ~~ ,
\end{eqnarray}
where $\Phi$ is gravitational (Bardeen) potential, first order in perturbation and $h_{ij}$'s are tensor perturbation of second order. 
\begin{figure}[H]
    \centering
    \includegraphics[height=8cm, width=12cm]{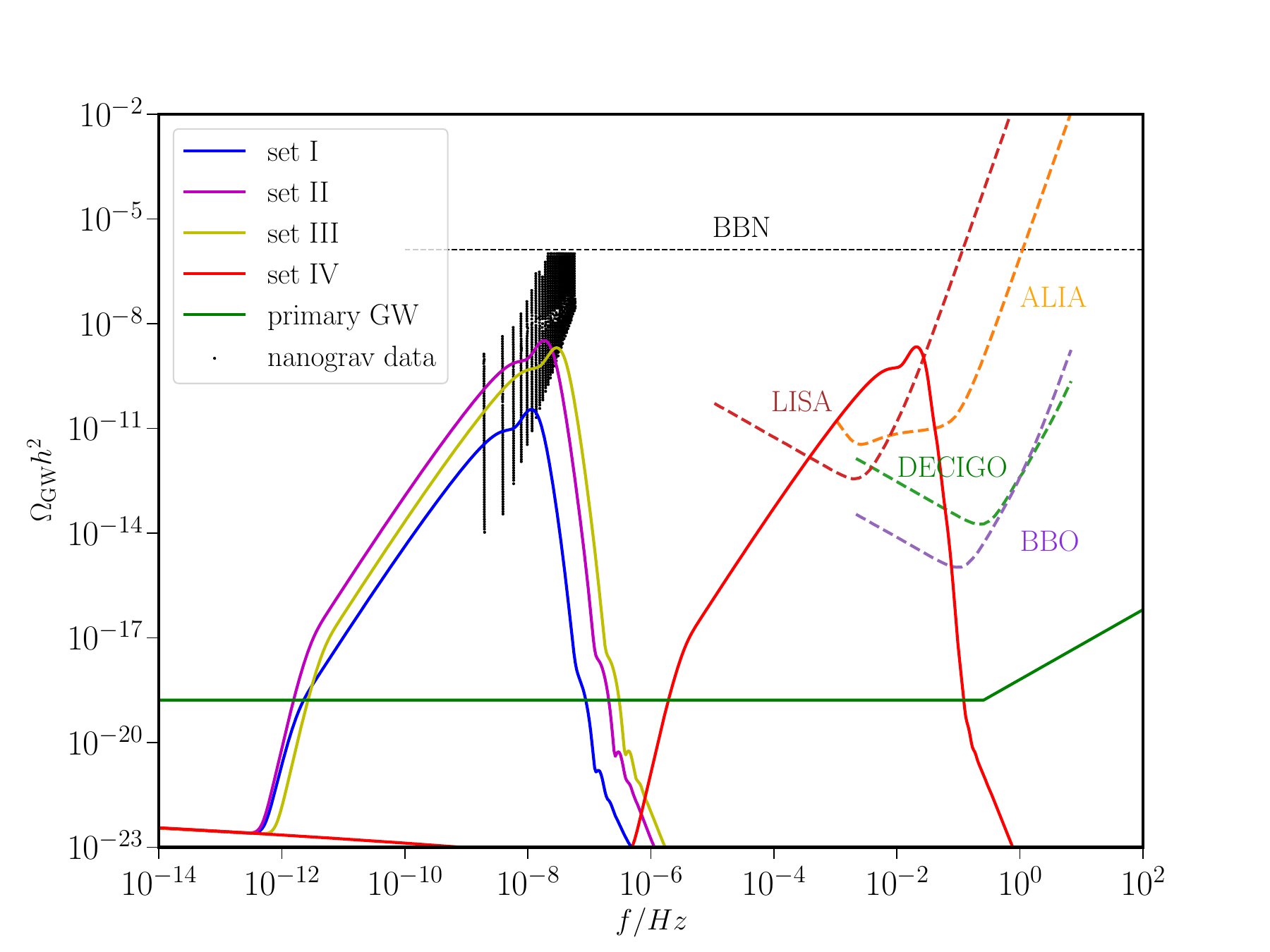}
    \caption{The figure depicts numerically obtained GW spectrum with frequency. The solid green line represents the primary spectra. The four bump like lines (blue, magenta, yellow and red) are IGWs spectra sourced by four different scalar spectra whose parameter space is given in table~\ref{table:parameters}. We have maintained same color code for IGWs and and its source-primary spectra. Sensitivity curve for future observations DECIGO, LISA, ALIA and BBO are also shown. The figure clearly shows that the IGWs falls into the region of NANOGrav 15 signal (set-I, set-II, set-III). For other parameter space (set -IV) IGWs fall in the sensitivity range of the mentioned future detectors.}
    \label{fig:GW01}
\end{figure}
The scalar perturbations in the gravitational potential $\Phi$ appear as the source term in the equations of motion for tensors; in terms of Fourier component equations of motion can be written as \cite{Kohri:2018awv},
\begin{align}
h''_{\bf k}(\eta) + 2 \mathcal{H} h'_{\bf k}(\eta)+ k^2 h_{\bf k}(\eta) =& 4 S_{\bf k}(\eta)~~\, . \label{EOM_h2nd} 
\end{align}
Here,  $S_{\bf k}(\eta)$ is the source term arising from the scalar perturbations, 
\begin{eqnarray}
    S_{\textbf{k}}(\eta) = \int \frac{d^3q}{(2\pi)^{3/2}} e_{ij}(\textbf{k})q_iq_j\left[2\Phi_{\textbf{q}}(\eta)\Phi_{\textbf{k-q}}(\eta) + \frac{4}{3(1+w)}(\mathcal{H}^{-1}\Phi_{\textbf{q}}^{\prime}(\eta) + \Phi_{\textbf{q}})\left(\mathcal{H}^{-1}\Phi_{\textbf{k-q}}^{\prime} + \Phi_{\textbf{k-q}}(\eta)\right)\right]\, , 
    \label{eq:Sk}
\end{eqnarray}
where $w= \frac{p}{\rho}$, is the equation-of-state parameter for the background fluid with pressure density $p$ and energy density $\rho$ and $\Phi_{\bf k}$ is the Fourier component of the gravitational potential. The power spectrum is defined as two point correlation function of the Fourier mode $h_k$: 
\begin{eqnarray}
  \bigl\langle h_{\bf k}^\lambda (\eta) h_{\bf k'}^{\lambda'}(\eta) \bigr\rangle = \delta_{\lambda \lambda'}\frac{2 \pi^2}{k^3} \delta(k+k') \mathcal{P}_{h}(k, \eta)\, .
  \label{eq:Ten_PS}
\end{eqnarray}
To evaluate the tensor power spectrum, we need to evaluate the quantity, $\langle S_{\bf k}(\eta) S_{\bf k'}(\eta') \bigr\rangle$ as well.  Hence, time evolution for $S_{\bf k}$ or $ \Phi_{\bf k}$ is needed to be known. The equation of motion for $\Phi_{\bf k}$ follows from Einstein's equation, 
\begin{eqnarray}
    \Phi_k'' + 3\mathcal{H}(1+c_s^2) \Phi_k' + (2\mathcal{H}' + (1+3c_s^2)\mathcal{H}^2 + c_s^2k^2)\Phi_k = \frac{a^2}{2}\tau \delta S\, ,
\end{eqnarray}
where sound speed $c_s^2=\left(\frac{\delta p}{\delta \rho}\right)_S$ and $\delta S$ is the the entropy perturbation.  The pressure perturbation can be written as $\delta P = c_s^2 \delta \rho + \tau \delta S$. We assume that the entropy perturbation is absent, that is $\delta S=0$ and speed of sound is constant, that is $c_s^2=w$. These reduce the equation of motion to \cite{Kohri:2018awv} 
\begin{eqnarray}
\Phi_{\bf{k}}''(\eta) +\frac{6(1+w)}{(1+3w)\eta}\Phi_{\bf k}'(\eta) + w k^2 \Phi_{\bf k }(\eta)=0\, ,
\label{eq:eomgravpot}
\end{eqnarray}
which in turn governs the evolution of the source term in Eq.~\eqref{eq:Sk}. We split $\Phi_{\bf k}(\eta)$ into primordial part $\phi_{\bf k}$ (not to be confused with inflaton) and transfer function $\Phi(k\eta)$, such that $\Phi_k=\Phi(k\eta) \phi_{\bf k}$. Taking two point correlation of the primordial part relates to the curvature perturbation as follows:
\begin{eqnarray}
\langle \phi_{\bf k} \phi_{\bf k'} \rangle = \delta ({\bf k}+{\bf k}' ) \frac{2\pi^2}{k^3} \left( \frac{3+3w}{5+3w} \right)^2 \mathcal{P}_S (k)~~.
\label{eq:pzetaInduced}
\end{eqnarray}
Now, to find the solution for $h_{\bf k}(\eta)$, we apply Green's function method to Eq.~\eqref{EOM_h2nd} as
\begin{eqnarray}
a(\eta) h_{\bf k}(\eta) = 4 \int^\eta \text{d}\bar{\eta} G_{\bf k}(\eta, \bar{\eta}) a(\bar{\eta}) S_{\bf k}(\bar{\eta})~~.
\label{eq:Ten_fluc}
\end{eqnarray}
where $G_{\bf k}(\eta, \bar{\eta})$ is the solution to the equation
\begin{eqnarray}
    G_{\bf k}''(\eta, \bar{\eta}) + \left(k^2 - \frac{a''(\eta)}{a(\eta)}\right)G_{\bf k}(\eta, \bar{\eta}) = \delta(\eta - \bar{\eta})\, .
\end{eqnarray}
Putting the solution to Eq.~\eqref{eq:Ten_fluc}, into  Eq.~\eqref{eq:Ten_PS}, we get \cite{Baumann:2007zm}, 
\begin{eqnarray}
  \langle h_{\bf k} (\eta)h_{\bf k'}(\eta') \rangle = \frac{16}{a^2(\eta)}\int_{\eta_{0}}^\eta d\bar{\eta_2} \int_{\eta_{0}} ^{\eta} d\bar{\eta_1} a(\bar{\eta_1})a(\bar{\eta_2})G_{\bf k}(\eta, \bar{\eta_1})G_{\bf k'}(\eta, \bar{\eta_2})\langle S_{\bf k} (\eta)S_{\bf k'}(\eta') \rangle\, .
  \label{eq:twopthk}
\end{eqnarray}
With the assumption that the curvature power spectrum is Gaussian, we compare Eq.~\eqref{eq:Ten_PS} with Eq.~\eqref{eq:twopthk}, to finally obtain the tensor power spectrum as
\cite{Kohri:2018awv, Baumann:2007zm}:
\begin{eqnarray}
\overline{\mathcal{P}_{h}(\eta,k)} &=& 4 
 \int_0^\infty \text{d}v \int_{\left| 1-v \right |}^{1+v}\text{d} u \left( \frac{4v^2 - (1+v^2-u^2)^2}{4vu} \right)^2 \nonumber \\
 &\times &\bar{I}{^2} (v,u,x) \mathcal{P}_S ( k v ) \mathcal{P}_S ( k u ), \label{eq:P_hsecond}\, 
\end{eqnarray}
where $x=k \eta $ and $\overline{\mathcal{P}_{h}(\eta,k)}$ is the time average of the dimensionless tensor power spectra.. The function $I$ is integration kernel; its full expression can be found in \cite{Baumann:2007zm,Kohri:2018awv,Espinosa:2018eve}. In the late time limit, $x\rightarrow \infty$, for modes entering horizon during radiation era, we have \cite{Kohri:2018awv} 
\begin{eqnarray}
   \bar{I^2_{R}} (v,u,x)= \frac{9}{32} \frac{(u^2+v^2 -3)^2}{u^6v^6 x^2}\left[ \pi^2 (u^2+v^2-3)^2 \Theta (u+v-\sqrt{3}) + \left( -4 u v + (u^2 + v^2-3)\ln\bigg \lvert \frac{3-(u+v)^2}{3-(u-v)^2} \bigg\rvert\right)^2 \right]\, ,
   \label{eq:Irad}
\end{eqnarray}
where $\Theta (x)$ is the Heaviside step function. 

Now we proceed to compute the energy density, which is defined in similar manner as that of the primary spectra in Eq.~\eqref{eq:GWprimary},
\begin{eqnarray}
    \Omega_{\text{GW}}(\eta,k) = \frac{1}{\rho_{\text{tot}}(\eta)}\frac{d\rho_{\text{GW}}(\eta,k)}{d\ln k} = \frac{1}{24}\left(\frac{k}{a(\eta)H(\eta)}\right)^2\overline{\mathcal{P}_{h}(\eta,k)}\, .
    \label{eq:OmegaGwsecondary}
\end{eqnarray}
Note that this definition differs from Eq.~\eqref{eq:GWprimary} by a factor of $1/2$ as we have not explicitly mentioned transfer function here. 
Using the expression for $\bar{I}(v,u,x)$ from Eq.~\eqref{eq:Irad} and $\frac{k}{a(\eta)H(\eta)}=x$ (for radiation era), we numerically integrate Eq.~\eqref{eq:P_hsecond} to obtain $\overline{\mathcal{P}_h(k, \eta)}$. 
Finally,  the GW energy spectrum for today is given as
\begin{eqnarray}
    \Omega_{\text{GW},0}(k) = 0.39 \left(\frac{g_{*}(T_{c})}{106.75}\right)^{-1/3} \Omega_{r,0}\Omega_{\text{GW}}(\eta_c,k)\, , 
    \label{eq:GWsecond}
\end{eqnarray}
where the pre-factor accounts for the transfer function for radiation era \cite{Kohri:2018awv,Correa:2023whf}, and $\eta_c$ donates the conformal time at horizon re-entry. We have taken $g_*(T_c)=106.75$ \cite{Kolb:1990vq} in all cases for our analysis. $\Omega_{\rm {GW}}(\eta_c, k)$ is evaluated from Eq.~\eqref{eq:OmegaGwsecondary}.

Using Eq.~\eqref{eq:freq_k}, we plot the GW energy density spectrum with respect to its frequency for four set of parameters given in table \ref{table:parameters}. The result obtained is shown in Fig.~\ref{fig:GW01}. It should be noted that the amplitudes of the scalar spectra peak around particular k values ($k \simeq 10^7 \rm Mpc^{-1}$ for first three sets and $k \simeq 10^{13} \rm Mpc^{-1}$ for the fourth set) as can be seen in Fig.~\ref{fig:powerspectra}. These peaks contribute maximum to IGWs at around $f = 10^{-8}$ Hz for first three sets and $f = 10^{-2}$ Hz for the fourth set as evident from Fig.~\ref{fig:GW01}. Energy spectrum for other frequencies are significantly dominated by primary GW backgrounds. It is apparent from Fig.~\ref{fig:GW01} that IGWs for the first three sets fairly satisfy the present NANOGrav data. The fourth set holds the promise of being detected in future missions like the DECi-hertz Interferometer Gravitational wave Observatory (DECIGO) \cite{Kawamura:2011zz}, the Laser Interferometer Space Antenna (LISA) \cite{Audley:2017drz}, the Advanced Laser Interferometer Antenna (ALIA) \cite{Bender:2004vw} and the Big Bang Observer (BBO) \cite{Harry:2006fi}. All the IGWs for the four sets as well as the blue-tilted kinetic regime for primary GW are well within the BBN constraint. 

\section{Conclusions } \label{sec:conclusion}
%PTA collaborations have presented substantial evidence for the existence of  SGWBs at nHz frequencies. If this {\color{red} GW backgrounds} indeed originated from cosmic sources, it implies that PTAs could offer a unique opportunity to probe the physics during early universe. Consequently, PTAs can be utilized to test both the composition of the universe at the time of gravitational wave generation and the underlying physics that contributed to the production of these gravitational waves. 
In this article, we deal with the quintessential potential with a Gaussian bump on it which generates enhanced scalar power spectrum in the region of interest without violating the CMB bounds. The addition of the bump deviates the potential from slow-roll conditions near the bump momentarily and mimics the ultra-slow roll situation. We obtain this after a numerical solution of the Klein-Gordon equation which describes the background dynamics. Furthermore, due to the deviation from slow-roll nature, we solve the Mukhanov-Sasaki equation numerically without any approximations to obtain the scalar and tensor power spectra. The spectra follows the slow-roll behaviour near CMB scales as can be roughly seen from Fig.~\ref{fig:powerspectra}. % At the pivot scale, $k = 0.05 \rm Mpc^{-1}$, the amplitude of power spectrum is  $2.1 \times 10^{-9}$, which exactly matches with the bound given by Planck 2018 data on CMB.%, while the $n_s-r$ lies in the $2\sigma$ region as per Planck data. % make it near cmb scale have ranges .05 is pivot
We find that our primary spectra dominates over the secondary except for those regions with enhanced scalar power spectra. We find that our primary spectra easily satisfies the BBN constraints even in the blue tilted region caused by kinetic domination. The amplitude of IGWs are significant where the scalar spectra are enhanced. We find that this significant IGWs occur/re-enter the horizon during the radiation dominated regime for the choice of our parameter space. So the presence of the kinetic epoch or blue-tilt does not affect these modes. All four sets of IGWs are also well inside the BBN bound on relevant scales.  Most importantly, we find that three sets (blue, magenta and yellow) well satisfy the NANOGrav data while the fourth set (red) lie in the region of future observations like DECIGO, LISA, ALIA and BBO, see Fig.~\ref{fig:GW01}. If these future missions detect high-amplitude gravitational waves (GWs) of cosmological origin, the model we are discussing (set 4) will be crucial in explaining such events. Additionally, a primordial black hole within the relevant mass range might be produced, potentially serving as a component of dark matter; which we defer to our future investigations. It is important to note that one can also utilize  data from different PTAs, leading to enhanced constraints. Consequently, this approach allows for better exploration of the parameter space, resulting in a more accurate comparison with the data.

A comment is in order in view of the ongoing active debate on the impact of quantum loop correction on PBH formation in single-field inflation.
%In addition to our main findings, it is crucial to highlight the broader context of the active consideration surrounding the topic of large primordial fluctuations and their potential collapse into primordial black holes.
%\st{Last but not the least, we should mention that the topic of the large primordial fluctuations and their collapse into primordial black holes is under active consideration at present.}
Studies confirm that loop corrections to the power spectrum shift the location of the peak of the {\it sharp} scalar fluctuation to the high frequency region if inflation is to proceed as desired \cite{Kristiano:2022maq, Riotto:2023hoz, Choudhury:2023vuj, Choudhury:2023jlt, Choudhury:2023rks, Choudhury:2023hvf, Choudhury:2023kdb, Kristiano:2023scm, Riotto:2023gpm, Firouzjahi:2023ahg, Firouzjahi:2023aum, Franciolini:2023lgy, Tasinato:2023ukp, Motohashi:2023syh}. As a result, the significant enhancement in IGWs could be visible in the high frequency region and could be probed by observational missions like LISA and others. It is interesting to note that the Galileon field is unique in a sense that its action does not receive quantum corrections thanks to its non-renormalizability property leaving the tree level results intact \cite{Choudhury:2023hvf}. Consequently, one can have significant amplitude of IGWs in both the low and high frequency regimes similar to our results which might in general be valid in case of a smooth slow roll to ultra-slow roll transition\footnote{This could be mimicked by a sharp bump/dip considered here.}. Furthermore, we should bring to notice  to some of  the alternative interpretations for the same NANOGrav result such as \cite{Madge:2023cak, Ashoorioon:2022raz, Guo:2023hyp, Choudhury:2023kam, Basilakos:2023xof, Broadhurst:2023tus, Oikonomou:2023qfz, Oikonomou:2023bah, Cheung:2023ihl}.

\section{Acknowledgements}
N.J. is supported by the National Postdoctoral Fellowship of the Science and Engineering Research Board (SERB), Department of Science and Technology (DST), Government of India, File No. PDF/2021/004114. 
The work of MS is supported by Science and Engineering Research Board (SERB), DST, Government of India under the Grant Agreement number CRG/2022/004120 (Core Research Grant). MS is also partially supported by the Ministry of Education and Science of the Republic of Kazakhstan, Grant
No. 0118RK00935 and CAS President's International Fellowship Initiative (PIFI).
%%%%%%%%%%%%

\section*{Appendix:A} 
\label{appendix:A}
Inflation is known to be responsible for generating the primordial perturbations. To understand the nature of these perturbations, we employ perturbation theory. At the linear order, the line element for scalar perturbations can be written in Newtonian gauge as follows \cite{Baumann:2009ds, Kinney2009tasi} 
\begin{eqnarray}
    ds^{2} = -(1+2\Phi)dt^{2} + a^2(t)\left(1-2\Psi\right)\delta_{ij}dx^{i} dx^j,
    \label{eq:scalarmetric}
\end{eqnarray}
where $\Phi$ and $\Psi$ are the two independent functions describing the scalar perturbations. In the absence of anisotropic stress $\Phi = \Psi$.  Additionally, perturbations arise from the perturbed inflaton field $\phi(t,\vec{x})$:
\begin{eqnarray}
    \phi(t,\vec{x}) = \phi_{0}(t) + \delta \phi(t,\vec{x}).
\end{eqnarray}
To quantify the perturbations, we introduce a gauge invariant comoving curvature perturbation, $\mathcal{R} = \Psi + \frac{H}{\dot{\phi_0}}\delta \phi$, which takes into account the perturbation terms from both the metric as well as the field. Using this curvature perturbation $\mathcal{R}$, we can write the action for inflaton \cite{Baumann:2009ds} as
\begin{eqnarray}
    S = \int d\eta d^{3}x\sqrt{-g}\left[\frac{R}{2} + \frac{g^{\mu\nu}\partial_{\mu}\phi\partial_{\nu}\phi}{2} - V(\phi)] + S^{(2)}(\mathcal{R}^{2}) + \dots \right],
\end{eqnarray}
where $R$ is the Ricci scalar and $S^{(2)}(\mathcal{R}^{2})$ is the second order action term given by
\begin{eqnarray}
    S^{(2)}(\mathcal{R}^{2}) = \int d\eta d^{3}xz^{2}\left[\mathcal{R}'^{2} - (\partial_{i}\mathcal{R})^{2}\right],
    \label{eq:2ndorderaction}
\end{eqnarray}
where `$'$' denotes derivative with respect to conformal time, $\eta$. Extremizing the action in Eq.~\eqref{eq:2ndorderaction}, we arrive at the following second order differential equation:
\begin{eqnarray}
    \mathcal{R}'' + 2\frac{z'}{z}\mathcal{R}' - \partial_{i}\partial^{i}\mathcal{R} = 0.
\end{eqnarray}
Here, $z = a(\eta)\sqrt{2\epsilon_1} = -\frac{1}{\eta H}\sqrt{2\epsilon_1}$. By decomposing $\mathcal{R}$ into its Fourier mode, $\mathcal{R}(\eta,\vec{x}) = \int \frac{d^{3}k}{(2\pi)^{3/2}}\mathcal{R}_k(\eta)e^{i\vec{k}\vec{x}}$, the above equation takes a more tractable form of:
\begin{eqnarray}
    \mathcal{R}_{k}'' + 2\frac{z'}{z}\mathcal{R}_{k}' +k^{2}\mathcal{R}_{k} = 0.
    \label{eq:Rk}
\end{eqnarray}

To simplify the analysis further, we define the Mukhanov-Sasaki variable, $v_{k} = \mathcal{R}_{k} z$, which leads to the Mukhanov-Sasaki equation, 
\begin{eqnarray}
    v_{k}'' + \left[k^{2} - \frac{z''}{z}\right]v_{k} = 0.
    \label{eq:mukhanovsasaki}
\end{eqnarray}

Under slow-roll approximations, this equation can be analytically solved, yielding the solution for $v_{k}$ as:
\begin{eqnarray}
    v_{k} = \frac{\alpha_k}{\sqrt{2k}}\left(1-\frac{i}{k\eta}\right)e^{-ik\eta} + \frac{\beta_k}{\sqrt{2k}}\left(1+\frac{i}{k\eta}\right)e^{ik\eta},
\end{eqnarray}
where $\alpha_k$ and $\beta_k$ are Bogoliubov coefficients. Now implementing the Bunch-Davies initial conditions, which corresponds to perturbations being in their vacuum state, gives $\alpha_k = 1$ and $\beta_k = 0$. So, 
\begin{eqnarray}
    \mathcal{R}_{k} = \frac{e^{-ik\eta}}{\sqrt{2k}}\left(1-\frac{i}{k\eta}\right)\left(\frac{-\eta H}{\sqrt{2\epsilon_1}}\right).
\end{eqnarray}
Finally, we take the two point correlation of the scalar perturbation, 
$\mathcal{R}_{k}$ and get the slow-roll approximated analytical result for scalar power spectrum
\begin{eqnarray}
    \mathcal{P}_{S} \equiv \frac{k^3}{2\pi^2}|\mathcal{R}_{k}|^2\bigg|_{-\eta \rightarrow 0} \simeq \frac{H^2}{8\pi^2\epsilon_1}.
\end{eqnarray}


\begin{thebibliography}{99}
%\cite{Guth:1980zm}
\bibitem{Guth:1980zm}
A.~H.~Guth,
%``The Inflationary Universe: A Possible Solution to the Horizon and Flatness Problems,''
Phys. Rev. D \textbf{23}, 347-356 (1981).
%doi:10.1103/PhysRevD.23.347
%9601 citations counted in INSPIRE as of 20 Jul 2023

\bibitem{starobinsky}
 A.~A.~Starobinsky, Phys. \ Lett.\ B {\bf 91}, 99 (1980).

%\cite{Linde:1981mu}
\bibitem{Linde:1981mu}
A.~D.~Linde,
%``A New Inflationary Universe Scenario: A Possible Solution of the Horizon, Flatness, Homogeneity, Isotropy and Primordial Monopole Problems,''
Phys. Lett. B \textbf{108}, 389-393 (1982).
%doi:10.1016/0370-2693(82)91219-9
%6082 citations counted in INSPIRE as of 20 Jul 2023
%%%.................NANOGRAV ........and othres ........NANOGRAV ...........NANOGRAV........ NANOGRAV......
\bibitem{NANOGrav:2023gor}
G.~Agazie \textit{et al.} [NANOGrav],
%``The NANOGrav 15 yr Data Set: Evidence for a Gravitational-wave Background,''
Astrophys. J. Lett. \textbf{951}, no.1, L8 (2023).
%\cite{Vagnozzi:2023lwo}
\bibitem{NANOGrav:2023hde}
G.~Agazie \textit{et al.} [NANOGrav],
%``The NANOGrav 15 yr Data Set: Observations and Timing of 68 Millisecond Pulsars,''
Astrophys. J. Lett. \textbf{951}, no.1, L9 (2023).

%\bibitem{NANOGrav:2023hfp}
%G.~Agazie \textit{et al.} [NANOGrav],
%``The NANOGrav 15-year Data Set: Constraints on Supermassive Black Hole Binaries from the Gravitational Wave Background,''
%[arXiv:2306.16220 [astro-ph.HE]].
%40 citations counted in INSPIRE as of 20 Jul 2023
%\cite{NANOGrav:2023tcn}
%\bibitem{NANOGrav:2023tcn}
%G.~Agazie \textit{et al.} [NANOGrav],
%``The NANOGrav 15-year Data Set: Search for Anisotropy in the Gravitational-Wave Background,''
%[arXiv:2306.16221 [astro-ph.HE]].
%8 citations counted in INSPIRE as of 20 Jul 2023
%\cite{NANOGrav:2023pdq}
%\bibitem{NANOGrav:2023pdq}
%G.~Agazie \textit{et al.} [NANOGrav],
%``The NANOGrav 15-year Data Set: Bayesian Limits on Gravitational Waves from Individual Supermassive Black Hole Binaries,''
%[arXiv:2306.16222 [astro-ph.HE]].
%12 citations counted in INSPIRE as of 20 Jul 2023
%\cite{NANOGrav:2023icp}
%\bibitem{NANOGrav:2023icp}
%A.~D.~Johnson \textit{et al.} [NANOGrav],
%``The NANOGrav 15-year Gravitational-Wave Background Analysis Pipeline,''
%[arXiv:2306.16223 [astro-ph.HE]].

\bibitem{Antoniadis:2023rey}
J.~Antoniadis, \textit{et al.} [EPTA+InPTA]
%``The second data release from the European Pulsar Timing Array III. Search for gravitational wave signals,''
[arXiv:2306.16214 [astro-ph.HE]].
\bibitem{Antoniadis:2023bjw}
J.~Antoniadis, \textit{et al.} [EPTA+InPTA],
%``The second data release from the European Pulsar Timing Array IV. Search for continuous gravitational wave signals,''
[arXiv:2306.16226 [astro-ph.HE]].
%13 citations counted in INSPIRE as of 22 Jul 2023
%\cite{Antoniadis:2023zhi}
\bibitem{Antoniadis:2023zhi}
J.~Antoniadis, \textit{et al.} [EPTA+InPTA],
%``The second data release from the European Pulsar Timing Array: V. Implications for massive black holes, dark matter and the early Universe,''
[arXiv:2306.16227 [astro-ph.CO]].
%41 citations counted in INSPIRE as of 22 Jul 2023
%\cite{Zic:2023gta}
\bibitem{Zic:2023gta}
A.~Zic, \textit{et al.} [PPTA],
%``The Parkes Pulsar Timing Array Third Data Release,''
[arXiv:2306.16230 [astro-ph.HE]].
%26 citations counted in INSPIRE as of 22 Jul 2023
%\cite{Reardon:2023zen}
\bibitem{Reardon:2023zen}
D.~J.~Reardon, \textit{et al.} [PPTA],
%``The Gravitational-wave Background Null Hypothesis: Characterizing Noise in Millisecond Pulsar Arrival Times with the Parkes Pulsar Timing Array,''
Astrophys. J. Lett. \textbf{951}, no.1, L7 (2023).
%doi:10.3847/2041-8213/acdd03
%[arXiv:2306.16229 [astro-ph.HE]].
%\cite{Reardon:2023gzh}
\bibitem{Reardon:2023gzh}
D.~J.~Reardon, \textit{et al.} [PPTA],
%``Search for an Isotropic Gravitational-wave Background with the Parkes Pulsar Timing Array,''
Astrophys. J. Lett. \textbf{951}, no.1, L6 (2023).
%doi:10.3847/2041-8213/acdd02
%[arXiv:2306.16215 [astro-ph.HE]].
%\cite{Xu:2023wog}
\bibitem{Xu:2023wog}
H.~Xu, \textit{et al.} [CPTA],
%``Searching for the Nano-Hertz Stochastic Gravitational Wave Background with the Chinese Pulsar Timing Array Data Release I,''
Res. Astron. Astrophys. \textbf{23}, no.7, 075024 (2023).
%doi:10.1088/1674-4527/acdfa5
%[arXiv:2306.16216 [astro-ph.HE]].
\bibitem{NANOGrav:2023hvm}
A.~Afzal \textit{et al.} [NANOGrav],
%``The NANOGrav 15 yr Data Set: Search for Signals from New Physics,''
Astrophys. J. Lett. \textbf{951}, no.1, L11 (2023).


%% NANO IGWS ...... NANO IGWS......  NANO IGWS  ....NANO IGWS ..... NANO IGWS 


%\cite{Ragavendra:2021qdu}

%doi:10.1103/PhysRevD.105.063533
%[arXiv:2108.04193 [astro-ph.CO]].
%18 citations counted in INSPIRE as of 23 Aug 2023


\bibitem{Franciolini:2023pbf}
G.~Franciolini, A.~Iovino, Junior., V.~Vaskonen and H.~Veermae,
%``The recent gravitational wave observation by pulsar timing arrays and primordial black holes: the importance of non-gaussianities,''
[arXiv:2306.17149 [astro-ph.CO]].


%\cite{Inomata:2023zup}
\bibitem{Inomata:2023zup}
K.~Inomata, K.~Kohri and T.~Terada,
%``The Detected Stochastic Gravitational Waves and Sub-Solar Primordial Black Holes,''
[arXiv:2306.17834 [astro-ph.CO]].

\bibitem{Cai:2023dls}
Y.~F.~Cai, X.~C.~He, X.~Ma, S.~F.~Yan and G.~W.~Yuan,
%``Limits on scalar-induced gravitational waves from the stochastic background by pulsar timing array observations,''
[arXiv:2306.17822 [gr-qc]].
\bibitem{Wang:2023ost}
S.~Wang, Z.~C.~Zhao, J.~P.~Li and Q.~H.~Zhu,
%``Exploring the Implications of 2023 Pulsar Timing Array Datasets for Scalar-Induced Gravitational Waves and Primordial Black Holes,''
[arXiv:2307.00572 [astro-ph.CO]].

\bibitem{You:2023rmn}
Z.~Q.~You, Z.~Yi and Y.~Wu,
%``Constraints on primordial curvature power spectrum with pulsar timing arrays,''
[arXiv:2307.04419 [gr-qc]].
\bibitem{Yi:2023mbm}
Z.~Yi, Q.~Gao, Y.~Gong, Y.~Wang and F.~Zhang,
%``The waveform of the scalar induced gravitational waves in light of Pulsar Timing Array data,''
[arXiv:2307.02467 [gr-qc]].
%\bibitem{Li:2023qua}
%J.~P.~Li, S.~Wang, Z.~C.~Zhao and K.~Kohri,
%``Primordial Non-Gaussianity and Anisotropies in Gravitational Waves induced by Scalar Perturbations,''
%[arXiv:2305.19950 [astro-ph.CO]].

\bibitem{Liu:2023ymk}
L.~Liu, Z.~C.~Chen and Q.~G.~Huang,
%``Implications for the non-Gaussianity of curvature perturbation from pulsar timing arrays,''
[arXiv:2307.01102 [astro-ph.CO]].
\bibitem{Balaji:2023ehk}
S.~Balaji, G.~Dom\`enech and G.~Franciolini,
%``Scalar-induced gravitational wave interpretation of PTA data: the role of scalar fluctuation propagation speed,''
[arXiv:2307.08552 [gr-qc]].

\bibitem{Figueroa:2023zhu}
D.~G.~Figueroa, M.~Pieroni, A.~Ricciardone and P.~Simakachorn,
%``Cosmological Background Interpretation of Pulsar Timing Array Data,''
[arXiv:2307.02399 [astro-ph.CO]].
%\cite{Balaji:2023ehk}
\bibitem{Bhattacharya:2020lhc}
S.~Bhattacharya, S.~Mohanty and P.~Parashari,
%``Implications of the NANOGrav result on primordial gravitational waves in nonstandard cosmologies,''
Phys. Rev. D \textbf{103}, no.6, 063532 (2021).



%%%%%%%%%%%%%%%%%%%%%%%%%%% REHEATING %%%%%%%%%%%%%%%%%%%%%%%%%%
\bibitem{Kofman:1994rk}
L.~Kofman, A.~D.~Linde and A.~A.~Starobinsky,
%``Reheating after inflation,''
Phys. Rev. Lett. \textbf{73}, 3195-3198 (1994).

\bibitem{Allahverdi:2010xz}
R.~Allahverdi, R.~Brandenberger, F.~Y.~Cyr-Racine and A.~Mazumdar,
%``Reheating in Inflationary Cosmology: Theory and Applications,''
Ann. Rev. Nucl. Part. Sci. \textbf{60}, 27-51 (2010).
\bibitem{Lozanov:2019jxc}
K.~D.~Lozanov,
%``Lectures on Reheating after Inflation,''
[arXiv:1907.04402 [astro-ph.CO]]

%%%%%%%%%%%%%%%%%%%%%%%%%%%%%%%%%%%%%%%%%%%%%%%%%%%
\bibitem{Ratra:1987rm}
B.~Ratra and P.~J.~E.~Peebles,
%``Cosmological Consequences of a Rolling Homogeneous Scalar Field,''
Phys. Rev. D \textbf{37}, 3406 (1988)
\bibitem{Peebles:1998qn}
P.~J.~E.~Peebles and A.~Vilenkin,
%``Quintessential inflation,''
Phys. Rev. D \textbf{59}, 063505 (1999).
%[arXiv:astro-ph/9810509 [astro-ph]].
%\cite{Sahni:2001qp}
\bibitem{Sahni:2001qp}
V.~Sahni, M.~Sami and T.~Souradeep,
%``Relic gravity waves from brane world inflation,''
Phys. Rev. D \textbf{65}, 023518 (2002).
%doi:10.1103/PhysRevD.65.023518
%[arXiv:gr-qc/0105121 [gr-qc]].
%224 citations counted in INSPIRE as of 20 Jul 2023


\bibitem{Sami:2004ic} 
  M.~Sami and N.~Dadhich,
  %``Unifying brane world inflation with quintessence'',
  TSPU Bulletin {\bf no. 7 (44)}, 25 (2004).
%  [hep-th/0405016].
  
\bibitem{Hossain:2014zma} 
  M.~W.~Hossain, R.~Myrzakulov, M.~Sami and E.~N.~Saridakis,
  %``Unification of inflation and dark energy à la quintessential inflation'',
  Int.\ J.\ Mod.\ Phys.\ D {\bf 24}, no. 05, 1530014 (2015).

\bibitem{Peebles:1999fz} 
  P.~J.~E.~Peebles and A.~Vilenkin,
  %``Noninteracting dark matter'',
  Phys.\ Rev.\ D {\bf 60}, no. 10, 103506 (1999).

\bibitem{Peloso:1999dm} 
  M.~Peloso and F.~Rosati,
  %``On the construction of quintessential inflation models'',
  JHEP {\bf 12}, 026 (1999).
%  [hep-ph/9908271].
\bibitem{Copeland:2000hn} 
  E.~J.~Copeland, A.~R.~Liddle and J.~E.~Lidsey,
  %``Steep inflation: Ending brane world inflation by gravitational particle production'',
  Phys.\ Rev.\ D {\bf 64}, no. 2, 023509 (2001).
\bibitem{Dimopoulos:2000md} 
  K.~Dimopoulos,
  %``Towards a model of quintessential inflation'',
  Nucl.\ Phys.\ Proc.\ Suppl.\ {\bf 95}, 70 (2001).
%  [astro-ph/0012298].


\bibitem{Majumdar:2001mm} 
  A.~S.~Majumdar,
  %``From brane assisted inflation to quintessence through a single scalar field'',
  Phys.\ Rev.\ D {\bf 64}, no. 8, 083503 (2001).
%  [astro-ph/0105518].

\bibitem{Rosenfeld:2005mt} 
  R.~Rosenfeld and J.~A.~Frieman,
  %``A simple model for quintessential inflation'',
  JCAP {\bf 09}, 003 (2005).
%  [astro-ph/0504191].
 \bibitem{Sami:2004xk} 
  M.~Sami and V.~Sahni,
  %``Quintessential inflation on the brane and the relic gravitational wave background,''
  Phys.\ Rev.\ D {\bf 70}, no. 8, 083513 (2004).
%  [hep-th/0402086].

  \bibitem{Tashiro:2003qp} 
  H.~Tashiro, T.~Chiba and M.~Sasaki,
  %``Reheating after quintessential inflation and gravitational waves,''
  Class.\ Quant.\ Grav.\  {\bf 21}, 1761 (2004).
  % doi:10.1088/0264-9381/21/7/004
%  [gr-qc/0307068].
\bibitem{Dimopoulos:2001ix} 
  K.~Dimopoulos and J.~W.~F.~Valle,
  %``Modeling quintessential inflation'',
  Astropart.\ Phys.\ {\bf 18}, 287 (2002).
%  [astro-ph/0111417].
\bibitem{WaliHossain:2014usl}
M.~Wali Hossain, R.~Myrzakulov, M.~Sami and E.~N.~Saridakis,
%``Unification of inflation and dark energy \`a la quintessential inflation,''
Int. J. Mod. Phys. D \textbf{24}, no.05, 1530014 (2015).
%doi:10.1142/S0218271815300141
%[arXiv:1410.6100 [gr-qc]].
\bibitem{Dimopoulos:2017zvq}
K.~Dimopoulos and C.~Owen,
%``Quintessential Inflation with $\alpha$-attractors,''
JCAP \textbf{06}, 027 (2017).
%[arXiv:1703.00305 [gr-qc]].
\bibitem{Giovannini:2003jw} 
  M.~Giovannini,
  %``Low-scale quintessential inflation'',
  Phys.\ Rev.\ D {\bf 67}, no. 12, 123512 (2003).
%  [hep-ph/0301264].
  
\bibitem{Tsujikawa:2013fta} 
  S.~Tsujikawa,
  %``Quintessence: A Review'',
  Class.\ Quant.\ Grav. {\bf 30}, no. 21, 214003 (2013).
%  [arXiv:1304.1961 [gr-qc]].

\bibitem{Hossain:2014xha} 
  M.~W.~Hossain, R.~Myrzakulov, M.~Sami and E.~N.~Saridakis,
  %``Variable gravity: A suitable framework for quintessential inflation'',
  Phys.\ Rev.\ D {\bf 90}, no. 2, 023512 (2014).
%  [arXiv:1402.6661 [gr-qc]].
 \bibitem{Ahmad:2019jbm}
S.~Ahmad, A.~De Felice, N.~Jaman, S.~Kuroyanagi and M.~Sami,
%``Baryogenesis in the paradigm of quintessential inflation,''
Phys. Rev. D \textbf{100}, no.10, 103525 (2019).
\bibitem{deHaro:2021swo}
J.~de Haro and L.~A.~Sal\'o,
%``A Review of Quintessential Inflation,''
Galaxies \textbf{9}, no.4, 73 (2021).
\bibitem{Bettoni:2021qfs}
D.~Bettoni and J.~Rubio,
%``Quintessential Inflation: A Tale of Emergent and Broken Symmetries,''
Galaxies \textbf{10}, no.1, 22 (2022).

\bibitem{Benisty:2020qta}
D.~Benisty and E.~I.~Guendelman,
%``Quintessential Inflation from Lorentzian Slow Roll,''
Eur. Phys. J. C \textbf{80}, no.6, 577 (2020)
Galaxies \textbf{10}, no.1, 22 (2022).

%doi:10.1088/1475-7516/2019/10/050
%[arXiv:1811.04093 [astro-ph.CO]].
%37 citations counted in INSPIRE as of 20 Jul 2023

%%INSTANT PREHEATING%%%%%%%%%%%%%INSTANT PREHEATING%%%%%%%%%INSTANT PREHEATING%%%%%%%
  \bibitem{Felder:1998vq} 
  G.~N.~Felder, L.~Kofman and A.~D.~Linde,
  %``Instant preheating,''
  Phys.\ Rev.\ D {\bf 59}, 123523 (1999).
\bibitem{Bassett:2005xm}
B.~A.~Bassett, S.~Tsujikawa and D.~Wands,
%``Inflation dynamics and reheating,''
Rev. Mod. Phys. \textbf{78}, 537-589 (2006).
%\cite{Kuroyanagi:2008ye}

%%%%%%%%BBN......................BBN............................BBN....
\bibitem{Figueroa:2018twl}
D.~G.~Figueroa and E.~H.~Tanin,
%``Inconsistency of an inflationary sector coupled only to Einstein gravity,''
JCAP \textbf{10}, 050 (2019).
\bibitem{Kuroyanagi:2008ye} 
  S.~Kuroyanagi, T.~Chiba and N.~Sugiyama,
  %``Precision calculations of the gravitational wave background spectrum from inflation,''
  Phys.\ Rev.\ D {\bf 79}, 103501 (2009).
%  [arXiv:0804.3249 [astro-ph]].
\bibitem{Ahmad:2017itq} 
  S.~Ahmad, R.~Myrzakulov and M.~Sami,
  %``Relic gravitational waves from quintessential inflation'',
  Phys.\ Rev.\ D {\bf 96}, no. 6, 063515 (2017).

\bibitem{Kuroyanagi:2014qaa} 
  S.~Kuroyanagi, S.~Tsujikawa, T.~Chiba and N.~Sugiyama,
  %``Implications of the B-mode Polarization Measurement for Direct Detection of Inflationary Gravitational Waves,''
  Phys.\ Rev.\ D {\bf 90}, no. 6, 063513 (2014).


  
\bibitem{Ferreira:1997hj}
P.~G.~Ferreira and M.~Joyce,
%``Cosmology with a primordial scaling field,''
Phys. Rev. D \textbf{58}, 023503 (1998).
\bibitem{Copeland:1997et}
E.~J.~Copeland, A.~R.~Liddle and D.~Wands,
%``Exponential potentials and cosmological scaling solutions,''
Phys. Rev. D \textbf{57}, 4686-4690 (1998).
\bibitem{Steinhardt:1999nw}
P.~J.~Steinhardt, L.~M.~Wang and I.~Zlatev,
%``Cosmological tracking solutions,''
Phys. Rev. D \textbf{59}, 123504 (1999).

%%%%tensor from scalar%%%%%%%%%%%%%%
\bibitem{Matarrese:1992rp}
S.~Matarrese, O.~Pantano and D.~Saez,
%``A General relativistic approach to the nonlinear evolution of collisionless matter,''
Phys. Rev. D \textbf{47}, 1311-1323 (1993). 


\bibitem{Matarrese:1993zf}
S.~Matarrese, O.~Pantano and D.~Saez,
%``General relativistic dynamics of irrotational dust: Cosmological implications,''
Phys. Rev. Lett. \textbf{72}, 320-323 (1994). 
%[arXiv:astro-ph/0507632 [astro-ph]].
%[arXiv:1908.03742 [gr-qc]].

\bibitem{Matarrese:1997ay}
S.~Matarrese, S.~Mollerach and M.~Bruni,
%``Second order perturbations of the Einstein-de Sitter universe,''
Phys. Rev. D \textbf{58}, 043504 (1998).
\bibitem{Noh:2003yg}
H.~Noh and J.~c.~Hwang,
%``Second-order perturbations of the friedmann world model,''
 Phys. Rev. D \textbf{ 69}, 104011 (2004).

\bibitem{Carbone:2004iv}
C.~Carbone and S.~Matarrese,
%``A Unified treatment of cosmological perturbations from super-horizon to small scales,''
Phys. Rev. D \textbf{71}, 043508 (2005).
\bibitem{Nakamura:2004rm}
K.~Nakamura,
%``Second-order gauge invariant cosmological perturbation theory: Einstein equations in terms of gauge invariant variables,''
Prog. Theor. Phys. \textbf{117}, 17-74 (2007).

%INDUCED GWS%%%%%%INDUCED GWS%%%%%%%%%%%%%%%%%%%%%%INDUCED GWS%%%%%%%%%%%%%%%%INDUCED GWS%%%%%%INDUCED GWS%%%
 \bibitem{Mollerach:2003nq}
S.~Mollerach, D.~Harari and S.~Matarrese,
%``CMB polarization from secondary vector and tensor modes,''
Phys. Rev. D \textbf{69}, 063002 (2004).
 \bibitem{Ananda:2006af}
K.~N.~Ananda, C.~Clarkson and D.~Wands,
%``The Cosmological gravitational wave background from primordial density perturbations,''
Phys. Rev. D \textbf{75}, 123518 (2007).

\bibitem{Baumann:2007zm}
D.~Baumann, P.~J.~Steinhardt, K.~Takahashi and K.~Ichiki,
%``Gravitational Wave Spectrum Induced by Primordial Scalar Perturbations,''
Phys. Rev. D \textbf{76}, 084019 (2007).
%\cite{Choudhury:2013woa}
\bibitem{Alabidi:2012ex}
L.~Alabidi, K.~Kohri, M.~Sasaki and Y.~Sendouda,
%``Observable Spectra of Induced Gravitational Waves from Inflation,''
JCAP \textbf{09}, 017 (2012).
\bibitem{Alabidi:2013lya}
L.~Alabidi, K.~Kohri, M.~Sasaki and Y.~Sendouda,
%``Observable induced gravitational waves from an early matter phase,''
JCAP \textbf{05}, 033 (2013).
 \bibitem{Kohri:2018awv}
K.~Kohri and T.~Terada,
%``Semianalytic calculation of gravitational wave spectrum nonlinearly induced from primordial curvature perturbations,''
Phys. Rev. D \textbf{97}, no.12, 123532 (2018).
%[arXiv:1804.08577 [gr-qc]].
\bibitem{Choudhury:2013woa}
S.~Choudhury and A.~Mazumdar,
%``Primordial blackholes and gravitational waves for an inflection-point model of inflation,''
Phys. Lett. B \textbf{733}, 270 (2014).


%[arXiv:1307.5119 [astro-ph.CO]].
%39 citations counted in INSPIRE as of 13 Jun 2023
%
\bibitem{Domenech:2021ztg}
G.~Dom\`enech,
%``Scalar Induced Gravitational Waves Review,''
Universe \textbf{7}, no.11, 398 (2021).
\bibitem{Domenech:2019quo}
G.~Dom\`enech,
%``Induced gravitational waves in a general cosmological background,''
Int. J. Mod. Phys. D \textbf{29}, no.03, 2050028 (2020).
\bibitem{Domenech:2020kqm}
G.~Dom\`enech, S.~Pi and M.~Sasaki,
%``Induced gravitational waves as a probe of thermal history of the universe,''
JCAP \textbf{08}, 017 (2020).



\bibitem{Correa:2023whf}
M.~Correa, M.~R.~Gangopadhyay, N.~Jaman and G.~J.~Mathews,
%``Induced Gravitational Waves via Warm Natural Inflation,''
[arXiv:2306.09641 [astro-ph.CO]].



\bibitem{Espinosa:2018eve}
J.~R.~Espinosa, D.~Racco and A.~Riotto,
%``A Cosmological Signature of the SM Higgs Instability: Gravitational Waves,''
JCAP \textbf{09}, 012 (2018).
\bibitem{Ragavendra:2021qdu}
H.~V.~Ragavendra,
%``Accounting for scalar non-Gaussianity in secondary gravitational waves,''
Phys. Rev. D \textbf{105}, no.6, 063533 (2022).
\bibitem{Edgar:2010}
Edgar Bugaev and  Peter Klimai, 
Phys. Rev. D \textbf{81}, 023517 
%.................PBHS.........PBHS...............................PBS..........

\bibitem{Hawking:1971ei}
S.~Hawking,
%``Gravitationally collapsed objects of very low mass,''
Mon. Not. Roy. Astron. Soc. \textbf{152}, 75 (1971).
\bibitem{Carr:1974nx}
B.~J.~Carr and S.~W.~Hawking,
%``Black holes in the early Universe,''
Mon. Not. Roy. Astron. Soc. \textbf{168}, 399-415 (1974).

%\cite{HosseiniMansoori:2023mqh}
\bibitem{HosseiniMansoori:2023mqh}
S.~A.~Hosseini Mansoori, F.~Felegray, A.~Talebian and M.~Sami,
%``PBHs and GWs from $\mathbb{T}^2$-inflation and NANOGrav 15-year data,''
[arXiv:2307.06757 [astro-ph.CO]].
%2 citations counted in INSPIRE as of 24 Jul 2023

\bibitem{Carr:2016drx}
B.~Carr, F.~Kuhnel and M.~Sandstad,
%``Primordial Black Holes as Dark Matter,''
Phys. Rev. D \textbf{94}, no.8, 083504 (2016).
\bibitem{Baumann:2007yr}
D.~Baumann, P.~J.~Steinhardt and N.~Turok,
%``Primordial Black Hole Baryogenesis,''
[arXiv:hep-th/0703250 [hep-th]].
\bibitem{Fujita:2014hha}
T.~Fujita, M.~Kawasaki, K.~Harigaya and R.~Matsuda,
%``Baryon asymmetry, dark matter, and density perturbation from primordial black holes,''
Phys. Rev. D \textbf{89}, no.10, 103501 (2014).
\bibitem{Atal:2019cdz}
V.~Atal, J.~Garriga and A.~Marcos-Caballero,
%``Primordial black hole formation with non-Gaussian curvature perturbations,''
JCAP \textbf{09}, 073 (2019).
%[arXiv:1905.13202 [astro-ph.CO]].
\bibitem{Mishra:2019pzq}
S.~S.~Mishra and V.~Sahni,
%``Primordial Black Holes from a tiny bump/dip in the Inflaton potential,''
JCAP \textbf{04}, 007 (2020).


\bibitem{Baumann:2009ds}
D.~Baumann,
%``TASI Lectures on Inflation,''
[arXiv:0907.5424 [hep-th]].
\bibitem{Kinney2009tasi}
W.~H.~Kinney,
%TASI Lectures on Inflation
[arXiv:0902.1529 [astro-ph.CO]].

\bibitem{Planck:2018jri}
Y.~Akrami \textit{et al.} [Planck],
%``Planck 2018 results. X. Constraints on inflation,''
Astron. Astrophys. \textbf{641}, A10 (2020).
%[arXiv:1807.06211 [astro-ph.CO]].
%\bibitem{Planck:2018vyg}
%N.~Aghanim \textit{et al.} [Planck],
%``Planck 2018 results. VI. Cosmological parameters,''
%Astron. Astrophys. \textbf{641}, A6 (2020)
%[erratum: Astron. Astrophys. \textbf{652}, C4 (2021)].

%\cite{Geng:2015fla}
\bibitem{Geng:2015fla} 
  C.~Q.~Geng, M.~W.~Hossain, R.~Myrzakulov, M.~Sami and E.~N.~Saridakis,
  %``Quintessential inflation with canonical and noncanonical scalar fields and Planck 2015 results,''
  Phys.\ Rev.\ D {\bf 92}, no. 2, 023522 (2015).
  %%CITATION = doi:10.1103/PhysRevD.92.023522;%%
  %34 citations counted in INSPIRE as of 19 Feb 2019
 %\cite{Geng:2017mic}
\bibitem{Geng:2017mic}
C.~Q.~Geng, C.~C.~Lee, M.~Sami, E.~N.~Saridakis and A.~A.~Starobinsky,
%``Observational constraints on successful model of quintessential Inflation,''
JCAP \textbf{06}, 011 (2017).
\bibitem{AresteSalo:2020yxl}
L.~Areste Salo and J.~Haro,
%``Quintessential Inflation for Exponential Type Potentials: Scaling and Tracker Behavior,''
Eur. Phys. J. C \textbf{81}, no.2, 105 (2021).
%\cite{Skugoreva:2019blk}
\bibitem{Skugoreva:2019blk}
M.~A.~Skugoreva, M.~Sami and N.~Jaman,
%``Emergence of cosmological scaling behavior in the asymptotic regime,''
Phys. Rev. D \textbf{100}, no.4, 043512 (2019).
%doi:10.1103/PhysRevD.100.043512
%[arXiv:1901.06036 [gr-qc]].
%10 citations counted in INSPIRE as of 20 Jul 2023
\bibitem{Basak:2021cgk}
S.~Basak, S.~Bhattacharya, M.~R.~Gangopadhyay, N.~Jaman, R.~Rangarajan and M.~Sami,
%``The paradigm of warm quintessential inflation and spontaneous baryogenesis,''
JCAP \textbf{03}, no.03, 063 (2022).


\bibitem{Giovannini:1998bp} 
  M.~Giovannini,
  %``Gravitational waves constraints on postinflationary phases stiffer than radiation,''
  Phys.\ Rev.\ D {\bf 58}, 083504 (1998)
  % doi:10.1103/PhysRevD.58.083504
  
\bibitem{Giovannini:1999bh} 
  M.~Giovannini,
  %``Production and detection of relic gravitons in quintessential inflationary models,''
  Phys.\ Rev.\ D {\bf 60}, 123511 (1999).
  % doi:10.1103/PhysRevD.60.123511
%  [astro-ph/9903004].

  %\cite{Giovannini:1999qj}
\bibitem{Giovannini:1999qj} 
  M.~Giovannini,
  %``Spikes in the relic graviton background from quintessential inflation,''
  Class.\ Quant.\ Grav.\  {\bf 16}, 2905 (1999).
  % doi:10.1088/0264-9381/16/9/308
%  [hep-ph/9903263].

%\cite{Riazuelo:2000fc}
\bibitem{Riazuelo:2000fc} 
  A.~Riazuelo and J.~P.~Uzan,
  %``Quintessence and gravitational waves,''
  Phys.\ Rev.\ D {\bf 62}, 083506 (2000).
%  doi:10.1103/PhysRevD.62.083506
%  [astro-ph/0004156].

%\cite{Giovannini:2008tm}
\bibitem{Giovannini:2008tm} 
  M.~Giovannini,
  %``Thermal history of the plasma and high-frequency gravitons,''
  Class.\ Quant.\ Grav.\  {\bf 26}, 045004 (2009)
  % doi:10.1088/0264-9381/26/4/045004
%  [arXiv:0807.4317 [astro-ph]].

%\cite{Artymowski:2017pua}
\bibitem{Artymowski:2017pua} 
  M.~Artymowski, O.~Czerwinska, Z.~Lalak and M.~Lewicki,
  %``Gravitational wave signals and cosmological consequences of gravitational reheating,''
  JCAP {\bf 1804}, no. 04, 046 (2018).
%  doi:10.1088/1475-7516/2018/04/046
%  [arXiv:1711.08473 [astro-ph.CO]].

\bibitem{Jaman:2022bho}
N.~Jaman and M.~Sami,
%``What Is Needed of a Scalar Field If It Is to Unify Inflation and Late Time Acceleration?,''
Galaxies \textbf{10}, no.2, 51 (2022).

%\cite{Cyburt:2015mya}
\bibitem{Cyburt:2015mya} 
  R.~H.~Cyburt, B.~D.~Fields, K.~A.~Olive and T.~H.~Yeh,
  %``Big Bang Nucleosynthesis: 2015,''
  Rev.\ Mod.\ Phys.\  {\bf 88}, 015004 (2016).
%  [arXiv:1505.01076 [astro-ph.CO]].

\bibitem{Kolb:1990vq}
E.~W.~Kolb and M.~S.~Turner,
%``The Early Universe,''
Front. Phys. \textbf{69}, 1-547 (1990).

%doi:10.1103/PhysRevD.97.123532
%[arXiv:1804.08577 [gr-qc]].
%310 citations counted in INSPIRE as of 24 Jul 2023

%\cite{Kawamura:2006up}
%\bibitem{Kawamura:2006up}
%S.~Kawamura, \textit{et al.}
%``The Japanese space gravitational wave antenna DECIGO,''
%Class. Quant. Grav. \textbf{23}, S125-S132 (2006).
%doi:10.1088/0264-9381/23/8/S17

%\cite{Kawamura:2011zz}
\bibitem{Kawamura:2011zz} 
  S.~Kawamura {\it et al.},
  %``The Japanese space gravitational wave antenna: DECIGO,''
  Class.\ Quant.\ Grav.\  {\bf 28}, 094011 (2011).
%  doi:10.1088/0264-9381/28/9/094011

%\cite{Audley:2017drz}
\bibitem{Audley:2017drz} 
  H.~Audley {\it et al.} [LISA Collaboration],
  %``Laser Interferometer Space Antenna,''
  arXiv:1702.00786 [astro-ph.IM].

%\cite{Bender:2004vw}
\bibitem{Bender:2004vw}
P.~L.~Bender,
%``Additional astrophysical objectives for LISA follow-on missions,''
Class. Quant. Grav. \textbf{21}, S1203-S1208 (2004)
doi:10.1088/0264-9381/21/5/120

%\cite{TheVirgo:2014hva}
\bibitem{Harry:2006fi}
G.~M.~Harry, P.~Fritschel, D.~A.~Shaddock, W.~Folkner and E.~S.~Phinney,
%``Laser interferometry for the big bang observer,''
Class. Quant. Grav. \textbf{23}, 4887-4894 (2006)
[erratum: Class. Quant. Grav. \textbf{23}, 7361 (2006)].

%\cite{Kristiano:2022maq}
\bibitem{Kristiano:2022maq}
J.~Kristiano and J.~Yokoyama,
%``Ruling Out Primordial Black Hole Formation From Single-Field Inflation,''
[arXiv:2211.03395 [hep-th]].

%\cite{Riotto:2023hoz}
\bibitem{Riotto:2023hoz}
A.~Riotto,
%``The Primordial Black Hole Formation from Single-Field Inflation is Not Ruled Out,''
[arXiv:2301.00599 [astro-ph.CO]].

%\cite{Choudhury:2023vuj}
\bibitem{Choudhury:2023vuj}
S.~Choudhury, M.~R.~Gangopadhyay and M.~Sami,
%``No-go for the formation of heavy mass Primordial Black Holes in Single Field Inflation,''
[arXiv:2301.10000 [astro-ph.CO]].

%\cite{Choudhury:2023jlt}
\bibitem{Choudhury:2023jlt}
S.~Choudhury, S.~Panda and M.~Sami,
%``No-go for PBH formation in EFT of single field inflation,''
[arXiv:2302.05655 [astro-ph.CO]].

%\cite{Choudhury:2023rks}
\bibitem{Choudhury:2023rks}
S.~Choudhury, S.~Panda and M.~Sami,
%``Quantum loop effects on the power spectrum and constraints on primordial black holes,''
[arXiv:2303.06066 [astro-ph.CO]].

\bibitem{Choudhury:2023hvf}
S.~Choudhury, S.~Panda and M.~Sami,
%``Galileon inflation evades the no-go for PBH formation in the single-field framework,''
[arXiv:2304.04065 [astro-ph.CO]].

%\cite{Choudhury:2023kdb}
\bibitem{Choudhury:2023kdb}
S.~Choudhury, A.~Karde, S.~Panda and M.~Sami,
%``Primordial non-Gaussianity from ultra slow-roll Galileon inflation,''
[arXiv:2306.12334 [astro-ph.CO]].
%1 citations counted in INSPIRE as of 24 Jul 2023

%\cite{Kristiano:2023scm}
\bibitem{Kristiano:2023scm}
J.~Kristiano and J.~Yokoyama,
%``Response to criticism on ''Ruling Out Primordial Black Hole Formation From Single-Field Inflation'': A note on bispectrum and one-loop correction in single-field inflation with primordial black hole formation,''
[arXiv:2303.00341 [hep-th]].
%38 citations counted in INSPIRE as of 24 Jul 2023

%\cite{Riotto:2023gpm}
\bibitem{Riotto:2023gpm}
A.~Riotto,
%``The Primordial Black Hole Formation from Single-Field Inflation is Still Not Ruled Out,''
[arXiv:2303.01727 [astro-ph.CO]].

%\cite{Firouzjahi:2023ahg}
\bibitem{Firouzjahi:2023ahg}
H.~Firouzjahi and A.~Riotto,
%``Primordial Black Holes and Loops in Single-Field Inflation,''
[arXiv:2304.07801 [astro-ph.CO]].

%\cite{Firouzjahi:2023aum}
\bibitem{Firouzjahi:2023aum}
H.~Firouzjahi,
%``One-loop Corrections in Power Spectrum in Single Field Inflation,''
[arXiv:2303.12025 [astro-ph.CO]].

%\cite{Franciolini:2023lgy}
\bibitem{Franciolini:2023lgy}
G.~Franciolini, A.~Iovino, Junior., M.~Taoso and A.~Urbano,
%``One loop to rule them all: Perturbativity in the presence of ultra slow-roll dynamics,''
[arXiv:2305.03491 [astro-ph.CO]].
%21 citations counted in INSPIRE as of 24 Jul 2023

%\cite{Tasinato:2023ukp}
\bibitem{Tasinato:2023ukp}
G.~Tasinato,
%``A large $|\eta|$ approach to single field inflation,''
[arXiv:2305.11568 [hep-th]].
%11 citations counted in INSPIRE as of 24 Jul 2023

%\cite{Motohashi:2023syh}
\bibitem{Motohashi:2023syh}
H.~Motohashi and Y.~Tada,
%``Squeezed bispectrum and one-loop corrections in transient constant-roll inflation,''
[arXiv:2303.16035 [astro-ph.CO]].
%17 citations counted in INSPIRE as of 24 Jul 2023

%%%%%%%%%%%%%%%%%%%%%%%%%%%%%%%%%%%%%%%%%%%%%%%%%%%%%%%%%
%\cite{Madge:2023cak}
\bibitem{Madge:2023cak}
E.~Madge, E.~Morgante, C.~Puchades-Ib\'a\~nez, N.~Ramberg, W.~Ratzinger, S.~Schenk and P.~Schwaller,
%``Primordial gravitational waves in the nano-Hertz regime and PTA data -- towards solving the GW inverse problem,''
[arXiv:2306.14856 [hep-ph]].
%26 citations counted in INSPIRE as of 02 Aug 2023

%\cite{Ashoorioon:2022raz}
\bibitem{Ashoorioon:2022raz}
A.~Ashoorioon, K.~Rezazadeh and A.~Rostami,
%``NANOGrav signal from the end of inflation and the LIGO mass and heavier primordial black holes,''
Phys. Lett. B \textbf{835}, 137542 (2022).
%doi:10.1016/j.physletb.2022.137542
%[arXiv:2202.01131 [astro-ph.CO]].
%52 citations counted in INSPIRE as of 02 Aug 2023

%\cite{Guo:2023hyp}
\bibitem{Guo:2023hyp}
S.~Y.~Guo, M.~Khlopov, X.~Liu, L.~Wu, Y.~Wu and B.~Zhu,
%``Footprints of Axion-Like Particle in Pulsar Timing Array Data and JWST Observations,''
[arXiv:2306.17022 [hep-ph]].
%36 citations counted in INSPIRE as of 02 Aug 2023

%\cite{Choudhury:2023kam}
\bibitem{Choudhury:2023kam}
S.~Choudhury,
%``Single field inflation in the light of NANOGrav 15-year Data: Quintessential interpretation of blue tilted tensor spectrum through Non-Bunch Davies initial condition,''
[arXiv:2307.03249 [astro-ph.CO]].
%7 citations counted in INSPIRE as of 02 Aug 2023

%\cite{Basilakos:2023xof}
\bibitem{Basilakos:2023xof}
S.~Basilakos, D.~V.~Nanopoulos, T.~Papanikolaou, E.~N.~Saridakis and C.~Tzerefos,
%``Signatures of Superstring theory in NANOGrav,''
[arXiv:2307.08601 [hep-th]].
%2 citations counted in INSPIRE as of 02 Aug 2023

%\cite{Broadhurst:2023tus}
\bibitem{Broadhurst:2023tus}
T.~Broadhurst, C.~Chen, T.~Liu and K.~F.~Zheng,
%``Binary Supermassive Black Holes Orbiting Dark Matter Solitons: From the Dual AGN in UGC4211 to NanoHertz Gravitational Waves,''
[arXiv:2306.17821 [astro-ph.HE]].
%15 citations counted in INSPIRE as of 02 Aug 2023

%\cite{Oikonomou:2023qfz}
\bibitem{Oikonomou:2023qfz}
V.~K.~Oikonomou,
%``Flat Energy Spectrum of Primordial Gravitational Waves vs Peaks and
the NANOGrav 2023 Observation,''
[arXiv:2306.17351 [astro-ph.CO]].
%5 citations counted in INSPIRE as of 11 Jul 2023

%\cite{Oikonomou:2023bah}
\bibitem{Oikonomou:2023bah}
V.~K.~Oikonomou,
%``Effects of the axion through the Higgs portal on primordial
gravitational waves during the electroweak breaking,''
Phys. Rev. D \textbf{107} (2023) no.6, 064071.
%doi:10.1103/PhysRevD.107.064071
%[arXiv:2303.05889 [hep-ph]].
%0 citations counted in INSPIRE as of 20 Apr 2023

%\cite{Cheung:2023ihl}
\bibitem{Cheung:2023ihl}
K.~Cheung, C.~J.~Ouseph and P.~Y.~Tseng,
%``NANOGrav Signal and PBH from the Modified Higgs Inflation,''
[arXiv:2307.08046 [hep-ph]].
%3 citations counted in INSPIRE as of 02 Aug 2023
\end{thebibliography}
\end{document}